\documentclass[12pt]{article}

\usepackage{amssymb,amsmath,amsfonts,eurosym,geometry,ulem,graphicx,caption,color,sectsty,comment,footmisc,caption,natbib,pdflscape,subfigure,array,setspace,hyperref}
\usepackage{lettrine}
\usepackage{float}
\usepackage{amssymb}
\usepackage{graphicx}
\usepackage{lscape}
\usepackage{adjustbox}
\usepackage{natbib}
\usepackage{longtable}
\usepackage{everypage}
\usepackage{longtable}
\usepackage{rotating}
\usepackage{lscape}
\usepackage{upgreek}
\usepackage{pdflscape}

\usepackage{bm}
\setcitestyle{year,open={(},close={)}} 
\newcommand{\blue}[1]{{\color{blue} #1}}

\newcolumntype{L}[1]{>{\raggedright\let\newline\\arraybackslash\hspace{0pt}}m{#1}}
\newcolumntype{C}[1]{>{\centering\let\newline\\arraybackslash\hspace{0pt}}m{#1}}
\newcolumntype{R}[1]{>{\raggedleft\let\newline\\arraybackslash\hspace{0pt}}m{#1}}

\begin{document}

\begin{titlepage}
\title{Forecasting Cryptocurrencies Log-Returns: a LASSO-VAR and Sentiment Approach }
\author{ Milos Ciganovic\thanks{Department of Economics and Law - Sapienza University of Rome. milos.ciganovic@uniroma1.it } \ Federico D'Amario\thanks{Department of Economics and Law - Sapienza University of Rome. federico.damario@uniroma1.it } }
\date{\today}
\maketitle
\begin{abstract}
\noindent Cryptocurrencies have become a trendy topic recently, primarily due to their disruptive potential and reports of unprecedented returns. In addition, academics increasingly acknowledge the predictive power of Social Media in many fields and, more specifically, for financial markets and economics. In this paper, we leverage the predictive power of Twitter and Reddit sentiment together with Google Trends indexes and volume to forecast the log returns of ten cryptocurrencies. Specifically, we consider $Bitcoin$, $Ethereum$, $Tether$, $Binance Coin$, $Litecoin$, $Enjin Coin$, $Horizen$, $Namecoin$, $Peercoin$, and $Feathercoin$. We evaluate the performance of LASSO-VAR using daily data from January 2018 to January 2022. In a 30 days recursive forecast, we can retrieve the correct direction of the actual series more than 50\% of the time. We compare this result with the main benchmarks, and we see a 10\% improvement in Mean Directional Accuracy (MDA). The use of sentiment and attention
variables as predictors increase significantly the forecast accuracy in terms of MDA but not in terms of Root Mean Squared Errors. We perform a  Granger causality test using a post-double LASSO selection for high-dimensional VARs. Results show no ``causality" from Social Media sentiment to cryptocurrencies returns\\
\vspace{0in}
Natural Language Processing\\
\vspace{0in}

\bigskip
\end{abstract}
\setcounter{page}{0}
\thispagestyle{empty}
\end{titlepage}
\pagebreak \newpage

\spacing{1.35}

\section{Introduction} \label{sec:introduction}
Since the introduction of Bitcoin by \citet{nakamoto2008bitcoin}, cryptocurrencies have gradually become very popular among investors. In the last decade, the world witnessed an unforeseen growth of cryptocurrencies, both in terms of their market capitalization and the number of kinds of coins. Many reasons may justify this boom:
first of all, social media and journals reported unprecedented returns of cryptocurrencies which led many, professional investors and not, to enter this market. This mechanism has been naturally motivated by minimal global regulation, which brought people to primarily see cryptocurrencies as means of payment for illegal trades. Eventually, their gigantic returns stimulated an enthusiasm reminiscent of the Gold Rush in the western U.S.
Investing in cryptocurrencies can be done quickly by everyone downloading an app on their smartphone. This led youngsters to represent, relatively, the main investors in digital currencies\footnote{See https://www.investopedia.com/younger-generations-bullish-on-cryptocurrencies-5223563. }. Moreover, due to the relatively young age of the cryptocurrency market, traditional news outlets cannot follow events timely.\\ Because of the reasons mentioned above, social media can be defined as the primary source of information for cryptocurrency investors. Specifically, micro-blogging websites such as Twitter \footnote{See https://twitter.com/.} and Reddit \footnote{See https://www.reddit.com/.} are widely used sources for cryptocurrency information. 
Significant fluctuations in cryptocurrencies' prices and their high volatility resulted in significant risks associated with investment in crypto assets. This has led to heated discussions about their place and role in the modern economy (see, for example \citet{corbet2019cryptocurrencies}, \citet{catalini2020some}, \citet{halaburda2020microeconomics} and \citet{auer2021permissioned}). Therefore, the issue of developing appropriate methods and models for predicting prices for digital currencies is relevant both for the scientific community and financial analysts, investors, and traders. A crucial contribution in terms of modelling and forecasting cryptocurrencies' financial time series has been given by  \citet{catania2019forecasting} and \citet{catania2021forecasting}  who develop a dynamic model suitable for the complex dynamics of these series as well as compare several alternative of univariate and multivariate models for point and density forecasts. Furthermore, many studies (see \cite{hitam2018comparative}, \cite{sun2020novel},  \cite{miller2021univariate}) show how Machine Learning algorithms are extremely convenient in terms of computational time and accuracy when forecasting cryptocurrencies' time series.
Lastly, increasing literature highlights the importance of specific factors that shape cryptocurrencies' demand helping in forecasting their prices, returns, and volatility.
According to the efficient market hypothesis \cite{fama1970efficient}, market prices reflect all available information, thus the prediction of stock returns should not be possible. On the other hand, considerable empirical evidences (see, e.g.  \citet{daniel2002investor} for a comprehensive review)  show that investors' psychology drives the stock market. This led many researchers to adopt sentiment indexes to improve forecast accuracy. \citet{glenski2019improved} exploit the predictive power of social signals from multiple platforms (GitHub and Reddit) to forecast prices for three cryptocurrencies. They show that social signals reduce error when forecasting daily coin prices and that the language used in comments within the official communities on Reddit are the best predictors overall. \citet{kraaijeveld2020predictive} show that Twitter sentiment has predictive power for the returns of several cryptocurrencies. \citet{aslanidis2022link} and \citet{nasir2019forecasting} highlight the link of Google Trends with cryptocurrencies regarding their returns and volatility.\\
In this study, we analyze the impact of sentiment variables on cryptocurrency returns by using a novel dataset that combines a number of social media, search engine data, and volume. We apply a state-of-the-art sentiment classification technique to investigate whether sentiment measures contain predictive power for returns. To the best of our knowledge, similar sets of predictors have not been employed jointly previously. We account for the high dimensionality of the predictor variables by using a regularization technique known as the LASSO. This allows us to investigate (i) whether the variables constructed from our novel dataset can help to improve log-return forecasts using a VAR approach compared to the benchmark models; (ii) which data source and which type of sentiment or attention measure is most relevant in terms of Granger-causality in High-Dimensional VARs.
Our results show that, on average, LASSO-VAR performs better in terms of Mean Directional Accuracy (MDA) than benchmark models. Moreover, the use of sentiment and attention variables as predictors increase significantly the forecast accuracy. We do not find Granger causality from sentiment indexes to cryptocurrencies returns. We find, instead, Granger causality between all cryptocurrencies except Bitcoin, Tether, and Feathercoin and from returns to the bitcoin sentiment extracted from Twitter. \\
The remainder of the paper is organized as follows. Section two \ref{sec:data} deals with data collection, describing the data set, its sources, and strategies to construct sentiment indexes. Section three \ref{sec: methods} describes the modelling strategy, the estimation, forecasting methods, and the metrics used for the comparative evaluation of the out-of-sample model predictions. Section four \ref{sec:result} summarizes some selected results. Section five \ref{sec:conclusion} concludes.

\section{Data Collection} \label{sec:data}
This study relies on multiple data sources. First, we collect daily data of ten cryptocurrencies from January 2018 to January 2022. We provide for the same period google trends and sentiment indexes from Twitter and Reddit. We complete our dataset with volume for each cryptocurrency considered.
\subsection{Cryptocurrency Data}
The cryptocurrencies used are reported in table \ref{crypto table} ranked in terms of Market Capitalization (MC) as of January 2022. We get our data from finance.yahoo.com\footnote{See: https://it.finance.yahoo.com/criptocurrencies/.}.  

\begin{table}[h!]
\caption{\footnotesize 10 Cryptocurrencies and their symbols, market capitalization (MCs) and rankings of MCs (as of 31 January 2022).}
\begin{adjustbox}{max width=\textwidth,center}
\label{crypto table}
\begin{tabular}{llll}
\hline
Cryptocurrency & Symbol & MC                & Rank by MC \\ \hline
               &        &                   &            \\
Bitcoin        & BTC    & \$702,864,225,136 & 1          \\
Ethereum       & ETH    & \$295,905,148,931 & 2          \\
Tether         & USDT   & \$78,188,468,450  & 3          \\
Binance Coin   & BNB    & \$63,930,448,963  & 4          \\
Litecoin       & LTC    & \$7,479,561,631   & 21         \\
Enjin Coin     & ENJ    & \$1,434,490,287   & 60         \\
Horizen        & ZEN    & \$505,212,977     & 120        \\
Namecoin       & NMC    & \$24,472,607      & 733        \\
Peercoin       & PPC    & \$13,809,105      & 837        \\
Feathercoin    & FTC    & \$1,927,251       & 1515       \\\hline   
\end{tabular}
\end{adjustbox}
\end{table}
Many reasons brought us to choose this set of cryptocurrencies. First of all, cryptocurrencies emerge and disappear continually, while our selected ten currencies have been publicly-traded consecutively. Moreover, all the currencies chosen have been created with a defined purpose representing innovative projects which brought development, progress, or value to the blockchain technology that Bitcoin had implemented. On the other hand, our sample includes three tier currencies as in \citet{gandal2016can}. $Bitcoin$, $Ethereum$, $Tether$, $Binance Coin$, whose market capitalizations stay in the world's top five, are ``top-tier" cryptocurrencies. $Litecoin$, $Enjin Coin$, $Horizen$, representing ``middle cryptocurrencies" in market capitalization. $Namecoin$, $Peercoin$, and $Feathercoin$ are representative ``minor cryptocurrencies" according to market capitalization. We include Tether (USDT) in our sample for a specific reason. We know, indeed, that it is a blockchain-based cryptocurrency whose tokens in circulation are backed by an equivalent amount of U.S. dollars, making it a stablecoin with a price pegged to USD \$1.00, which leads this currency to be very low volatile. Tether was designed to build the necessary bridge between fiat currencies and cryptocurrencies and offer users stability, transparency, and minimal transaction charges. We decided to include it as a counterfactual to understand whether sentiment indexes can help the prediction of stablecoin currencies.
We compute the log returns and include them in our sample. Table \ref{summary table} provides several summary statistics.
\begin{table}[h!]
\centering
\captionsetup{font=footnotesize}
\caption{Log-returns summary statistics for the ten cryptocurrencies during the period 1 January 2018 to 31 January 2022}
\label{summary table}
\begin{adjustbox}{max width=\textwidth,center}
\begin{tabular}{lcccccccccc}
\hline
         & BTC-USD  & ETH-USD  & USDT-USD & BNB-USD  & LTC-USD  & ENJ-USD & ZEN-USD  & NMC-USD  & PPC-USD  & FTC-USD  \\ \hline
Mean     & 0.001    & 0.001    & 0        & 0.003    & 0        & 0.002   & 0        & -0.001   & -0.001   & -0.003   \\
Median   & 0.001    & 0.001    & 0        & 0.001    & 0        & -0.001  & -0.001   & 0.001    & -0.001   & -0.005   \\
Min      & -0.465   & -0.551   & -0.053   & -0.543   & -0.449   & -0.624  & -0.546   & -1.16    & -0.665   & -0.474   \\
Max      & 0.172    & 0.231    & 0.053    & 0.529    & 0.291    & 0.768   & 0.38     & 0.75     & 0.567    & 0.409    \\
Range    & 0.637    & 0.781    & 0.106    & 1.072    & 0.74     & 1.392   & 0.926    & 1.91     & 1.232    & 0.883    \\
Skew     & -1.147   & -1.099   & 0.3      & 0.3      & -0.608   & 1.132   & -0.233   & -0.869   & -0.174   & -0.174   \\
kurtosis & 14.042   & 10.795   & 34.34    & 15.579   & 7.947    & 15.508  & 6.268    & 19.999   & 13.593   & 4.575    \\
ADF      & -11.0753 & -11.1057 & -14.186  & -11.1936 & -11.3664 & -11.201 & -11.0829 & -12.9631 & -12.9892 & -11.5868 \\ \hline
\end{tabular}
\end{adjustbox}
\footnotesize {Notes: all ADF statistics are stationary at 1\% level}

\end{table}

\subsection{Google Trends Data}
We collected forty three google trends searches   (Table \ref{gtrends} in the Appendix reports the list of all the Google trends collected). 
\begin{figure}[H]
\caption{Google trends words collected}
\label{word}
\centering
     \includegraphics[scale = 0.6]{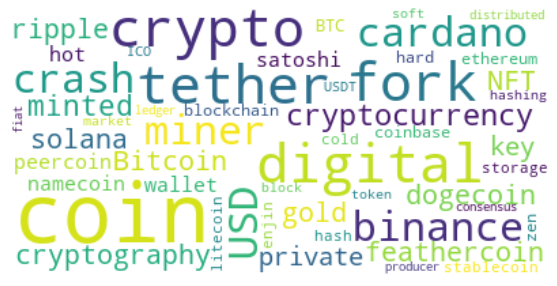}
     \label{fig:words}
 \end{figure} 
 Google Trends is a search trend feature that shows how frequently a given search term is entered into Google’s search engine relative to the site’s total search volume over a given period. Google Trends are available at a daily frequency only if the selected period is less than nine months. If the time frame is between nine months and five years, weekly data are provided; if it is longer than five years, data are monthly. A trivial solution like querying the data month by month and then tying it together will not work in this case because Google Trends assess interest in relative values within the given time. It means that for a given keyword and month, Google Trend will estimate the interest identically, with a local minimum of 0 and a local maximum of 100, for events in one month even if they had twice as many searches than in the other. To get proper daily estimates, we procede as follows:\\
1 - Query daily estimates for each month in the specified time frame;\\
2 - Queries monthly data for the whole time frame;\\
3 - Multiply daily estimates for each month from step 1 by its weight from step 2.

\subsection{Twitter and Reddit Sentiment Indexes}
 With regard to sentiment indexes, we use the Twitter API v2 \footnote{See https://developer.twitter.com/en/docs/twitter-api} to download tweets using the name of each cryptocurrency from our sample as a search parameter. We end up with 13,195,084 tweets from 2,253,469 unique users from January 2018 to January 2022. Each tweet's query provides the user with the timestamp (UTC+0) and the text containing a maximum of 280 characters. Concerning Reddit data, we use the Pushshift Reddit API \citet{baumgartner2020pushshift} to collect comments under the official subreddit of each cryptocurrency included in our sample. We gather a total of 4,406,897 comments from 420,024 unique users in the same time frame specified above. We are aware that increasing the daily quantity of tweets and comments collected would enhance the quality and results of our study. However, we believe that our sample is enough to get the representative social sentiment we are looking for. \\
 The sentiment time series are created using Valence Aware Dictionary and Sentiment Reasoner (VADER) algorithm \citet{hutto2014vader}. It uses a list of lexical features, optimized for social media texts, (words, punctuation, and emoticons) labelled as positive or negative according to their semantic orientation to calculate the text sentiment. Vader sentiment returns the probability of a given input sentence to be positive, negative, and neutral. In particular, every tweet and comment are passed to the algorithm. Therefore, we retrieve a Valence score for each of them which is measured on a scale from -4 to +4, where -4 stands for the most 'Negative' sentiment and +4 for the most 'Positive' sentiment. Midpoint 0 represents a 'Neutral' Sentiment.
 The unidimensional measure the algorithm can retrieve is called the ``Compound Score". It is computed by summing the valence scores of each word in the lexicon, adjusted according to the rules, and then normalized to be between -1 (most extreme negative) and +1 (most extreme positive):\\
\begin{equation}
    x = \frac{x}{\sqrt{x^2+\alpha}}
\end{equation}
Where $x$ is the sum of the valence score of constituent words, and $\alpha$ is a normalization constant (the default value is $15$).
Finally, we obtain the compound score for each input processed and we aggregate them into daily frequency by taking the average.

\subsection{Volume}
We complete our dataset with cryptocurrencies volume. The dynamic volume-return could bring valuable information which can be used for price predictions or trading strategies. Generally speaking, the literature concerning the relationship between volume and returns suggests that there is a positive correlation between them (see, e.g. \citet{jain1988dependence} and \citet{llorente2002dynamic}). Concerning cryptocurrencies, there is a growing literature studying the volume-return relation. \citet{balcilar2017can} employ a non-parametric causality-in-quantiles test to assess the causality relation between trading volumes and bitcoin returns and volatility. They show that volume can predict returns, except during bull and bear periods in the Bitcoin market. \citet{zhang2018multifractal} investigate the return-volume relationship of the Bitcoin market based on multifractal detrended cross-correlation. They found that Bitcoin exhibits a non-linear dependent relationship in return-volume with cross-correlations of return-volume showing anti-persistent behaviour. Another, \citet{naeem2020extreme} explore extreme return-volumes dependence among the highest capitalized cryptocurrencies using the Copula approach. They discovered that coefficients of lower tail dependence are significant for Bitcoin, Ripple, and Litecoin, which means that low volumes follow low returns. Lower tail dependence for the return-volume relationship is stronger than the upper tail dependence for Bitcoin, Ripple, and Litecoin. Moreover, for negative return-volume, left tail dependence coefficients are significant for Ripple and Litecoin, which means that low volumes follow high returns for Ripple and Litecoin. Finally, in a recent work, \citet{chan2022extreme} investigate the extreme dependence and correlation between high-frequency cryptocurrency returns and transaction volumes, at the extreme tails associated with booms and busts in the cryptocurrency markets. Applying an extreme value theory approach, they found a weak positive correlation between return and volume at the tails.

\section{Methods}
\label{sec: methods}
\subsection{LASSO-VAR}
The LASSO is a method for automatic variable selection and parameters shrinkage. It can be used to select the most informative predictors of a target variable Y from a set of variables and parameters, possibly larger than sample information, virtually making high-dimensional modelling and forecasting feasible for any degree of model dimension and complexity.
\noindent The LASSO has been initially developed for a single equation setting by  \citet{tibshirani1996regression}. The LASSO approaches curve fitting as a quadratic programming problem, where the objective function penalizes the total size of the regression coefficients based on the value of a tuning parameter, $\lambda$. In doing so, the LASSO can drive the coefficients of irrelevant variables to zero, thus performing the automatic variable selection.
The strength of the penalty must be tuned. The stronger the penalty, the higher the number of coefficients that are shrunk to zero. The model is thus forced to select only the most important predictors, i.e. those with the highest contribution to the prediction of the target variable.

\noindent Let $\{y_t\}_{t=1}^T$ be a $K$ dimensional multiple time series process generated by VAR process of order \textit{p}, denoted as VAR(\textit{p}):
\begin{equation} \label{VAR}
\begin{split}
 y_t = A_1 y_{t-1}& + \ldots + A_p y_{t-p} + u_t,\\
 &u_t \sim \mathcal{N}(0, \mathbf{\Sigma}_u).
\end{split}
\end{equation}
\newline
\noindent In our analysis, we fix the maximum order of lag \textit{p} to fourteen days, as suggested by the Bayesian Information Criterion, calculated over the full sample, i.e. including data up to the end of January 2022. 
 Notice that, since we standardize data before modelling, the \textit{K}-dimensional intercept vector is not considered in the VAR. Each $A_i$ is a $K \times K$ matrix of coefficients for the endogenous variables, and  $u_t\stackrel{\text{wn}}{\sim}(0,\mathbf{\Sigma}_u)$ is the vector of reduced-form errors.

\noindent For notation convenience we will introduce a compact matrix representation of \eqref{VAR}\\ 
\begin{equation*}
  \begin{split}
   \mathbf{Y} &= [y_{p+1},\ldots,y_T]\\
   \mathbf{Z}_t &= [y'_p,\ldots,y'_{t-p}]\\
   \mathbf{Z} &= [\mathbf{Z}_p;\ldots;\mathbf{Z}_{T-p}]\\
   \mathbf{A} &= [A_1,\ldots,A_p]\\
   \mathbf{U}&= [u_1,\ldots,u_T]
    \end{split}
\quad
  \begin{split}  
    &(K \times T);\\
    &[1 \times Kp];\\
    &(T \times Kp)';\\
    &(K \times Kp);\\
    &(K \times T)
    \end{split}
\end{equation*}
We can express the VAR as
\begin{equation}
    \mathbf{Y = AZ+U},
\end{equation}
with $\mathbf{U} \sim\mathcal{N}(0,\mathbf{I_t \otimes \Sigma_u})$.
\\The LASSO objective function is minimized as follows:
\begin{equation}\label{lasso}
\hat{\mathbf{A}}(\lambda) = \arg\min_{\mathbf{A}} \doublespacing  \frac{1}{T}\|\mathbf{AZ - Y }\|_{2}^2+\lambda\|\mathbf{A}\|_{1}\\,
\end{equation}

\noindent where $\lambda$ is the shrinkage parameter.

\noindent Together with the homoskedastic framework described above, we decide to account for heteroskedasticity estimating the model through Feasible GLS estimator assuming
an $AR(1)$ process for the errors. Then $\mathbf{\Omega} = (\mathbf{I_t \otimes \Sigma_u})$ will be a $T \times T$  Toeplix matrix of the auto-correlation structure of the data.

\noindent By applying the LASSO procedure, we seek to obtain a sparse structure for the coefficient matrices.
\noindent The optimization problem is solved by applying a coordinate descent numerical procedure, as explained in  \citet{kim2007interior} and  \citet{friedman2010regularization}. 

\subsection{Calibrating the LASSO-VAR through time series cross-validation}
As immediately evident from \eqref{lasso}, lambda ($\lambda$) is the most important parameter in the LASSO framework. The selection of the best predicting model depends on its calibration, which should not be sample-specific. A cross-validation stage is thus employed to get the ``optimal" value for $\lambda$. 

\noindent In this respect, we emply an expanding window (more precisely, an ``anchored walk forward") approach to cross-validation, which is the standard practice in many analyses considering time series modelling. In practice, the data is divided into a training and a test sample. The test set is being held for final evaluation, whereas the training set is further split into three subsets. 

\begin{figure}[H]
    \centering
    \includegraphics[width=7cm]{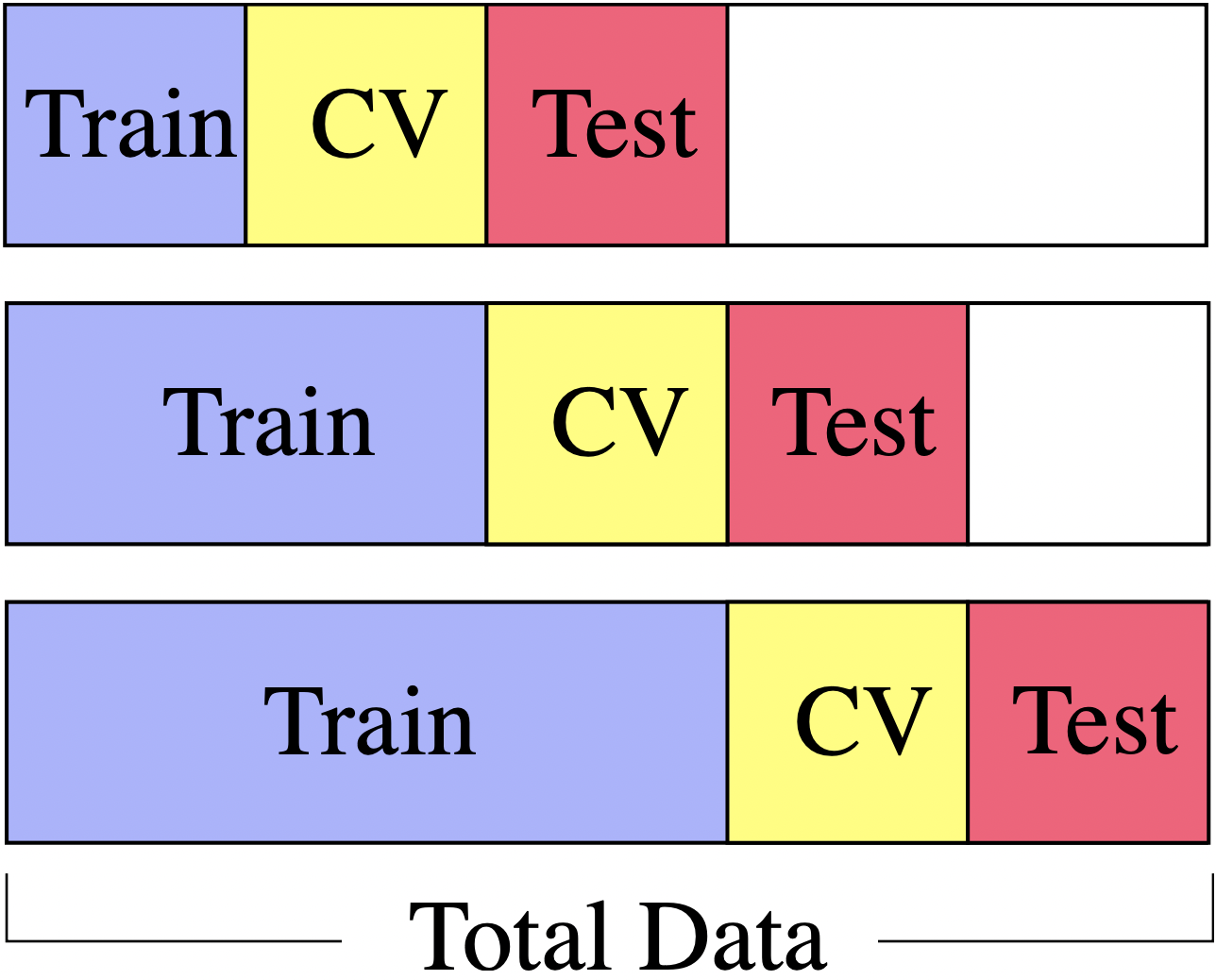}
    \caption{Graphical example of the 3-split time series cross validation}
    \label{tscv}
\end{figure}

\noindent An example of 3-split time series cross-validation method is depicted in Figure \ref{tscv}.

\noindent Time series cross-validation essentially takes into consideration time dependence, such that the training set is always built considering observations that occurred prior to the test set. The anchored walk forward cross-validation method implies a gradually expanding training set, pushing forward a fixed dimension test set.

\subsection{Evaluation of the forecasting performances}\label{evaluation}

After estimating the sparse model resulting from the cross-validated LASSO-VAR, we proceed to forecast. We use the sample from $01/01/2018–12/31/2021$ to obtain initial parameter estimates for all models, which are then used to predict outcomes from 01/01/2022 $(h = 1)$ to $01/04/2022$ $(h = 4)$. In the next period, we include data for $01/01/2022$ in the estimation sample and use the resulting estimates to predict the outcomes from $01/02/2022$ to $01/06/2022$. We proceed recursively in this fashion until $01/31/2022$, thus generating a time series of point forecasts.
In \eqref{VARforec} we show the standard procedure to obtain out-of-sample forecasts when a VAR is assumed to be the data generating process.
\begin{equation} \label{VARforec}
    y_{T+h \mid T} = c + A_1y_{T+h-1 \mid T}  + \ldots + A_py_{T+h-p \mid T}.
\end{equation}
\noindent In order to evaluate the forecasting model performances, we adopt the Root Mean Squared Error (RMSE) and the Mean Directional Accuracy (MDA):

\begin{equation}
    RMSE = \sqrt{\frac{1}{H}\sum_{\tau=\underline{t}}^{\bar{t}-h}e^2_{i,\tau+h}}\quad,
\end{equation}
\begin{equation}
    MDA = \frac{1}{H}\sum_{\tau=\underline{t}}^{\bar{t}-h}sign(y_{i,\tau+h}-y_{i,\tau+h-1})==sign(\hat{y}_{i,\tau+h}-\hat{y}_{i,\tau+h-1}).
\end{equation}

We evaluate the performances of our selected models by comparing the results with the two  multivariate high-dimensional time series forecasting model alternatives in discrete time. Specifically, we consider the Large Bayesian VAR (LBVAR) introduced by \citet{banbura2010large} and  Factor Augmented VAR (FAVAR) proposed by \citet{bernanke2005measuring}.\\
Finally, in order to test the statistical significance of differences in point forecasts, we consider pairwise tests of equal predictive accuracy (henceforth, EPA; \citet{diebold1995diebold}; \citet{west1996asymptotic}) in terms of RMSE. All EPA tests we conduct are based on a two sided test with the null hypothesis being the LBVAR and FAVAR benchmarks. We use standard normal critical values. Based on simulation evidence in \citet{clark2013advances}, when computing the variance estimator which enters the test statistic we rely on serial correlation robust standard errors and incorporate the finite sample correction due to \citet{harvey1997testing}. In Table \ref{RMSE}, we use \blue{***}, \blue{**} and \blue{*} to denote results which are significant at the 1\%, 5\% and 10\% levels, respectively, in favor of the model listed at the top of each column compared with LBVAR. We use {***},{ **} and {*} when we compare our models with FAVAR. \\

\subsection{High-dimensional Granger causality test}
The concept of Granger causality captures the predictability given a particular information set \citet{granger_1969_testing, granger1980testing}. If the addition of variable ${I}$ to the given information set $\Xi$ alters the conditional distribution of another variable $J$, and both $I$ and $\Xi$ are observed prior to $J$, then $I$ improves the predictability of $J$ and is said to Granger cause $J$ with respect to $\Xi$. \citet{granger_1969_testing} considers $\Xi$ in a theoretical and non-practicable way as all the information available in the universe. The choice of the information set plays thus a crucial role in determining  (non-)Granger causality. Cases of spurious Granger causality from $I$ to $J$ may arise when both $I$ and $J$ are Granger caused by $Q$, but $Q$ is omitted from $\Xi$. High-dimensional models may avoid possible spurious causality due to omitted variables. Of course, conditioning on so many variables leads to the curse of dimensionality, rendering many standard statistical techniques invalid. To overcome the problem mentioned above, regularized estimation procedures as the LASSO can be used. As described in \citet{hecq2019granger}, one might be tempted to perform the LASSO as in \eqref{lasso} on \eqref{wrongGC}\footnote{Where $y_J = vec(\mathbf{Y_J})$ denote the $N_J\times1$ stacked vector containing all the observations of the variables in $J$. Similarly $u_J = vec(\mathbf{U}_J)$, $Z^\otimes = \mathbf{I}_{N_J}\otimes\mathbf{Z}$ and $\beta = vec(\mathbf{A})$. $GC$ and $-GC$ stand respectively for chosen possible Granger causing variables and remaining variables.}:
\begin{equation}\label{wrongGC}
    y_J = Z^\otimes\beta + u_J = Z^\otimes_{GC}\beta_{GC}+Z^\otimes_{-GC}\beta_{-GC}+u_J,
\end{equation}
testing the parameter significance of the selected ``Granger-causing" variables. However, this procedure does not consider that the final and selected model is random and function of the data. The randomness in the selection step means that post-selection estimators do not converge uniformly to a normal distribution, as the potential omitted variable bias from omitting weak, but still relevant, variables in the selection step is too large to perform uniformly valid inference. Many authors tried to cope with this issue (see \citet{hecq2019granger} for a comprehensive review). We adopt the procedure proposed by \citet{hecq2019granger} who specifically implemented a post-double-selection procedure, initially developed by \citet{belloni2014high}, in a VAR context. The idea behind their approach is that $\Xi$ is made of variables of interest, possible Granger-causing variables and the remaining variables. Their idea is to regress both the variable of interest and possible Granger-causing variables on the remaining variables. These regressions could be high-dimensional and cannot be estimated by least square.
The remaining variables retained are those whose parameter is significant in at least one of the regressions mentioned above. In this case, the omitted variable bias will only occur if the LASSO fails to select a relevant variable in both regressions simultaneously. As the probability for this to occur decreases quadratically, this is negligible asymptotically and allows for valid inference.\\
In our application we test the (non-)Granger Causality for all the variables in our dataset considering all the possible combinations.  
Furthermore, since the time series length of our variables is much higher in magnitude compared to the number of covariates, we do not need to set any bound on the penalty to ensure that in each  regression a maximum of variables are selected. In this case, we tune $\lambda$ selecting the one which minimizes the Bayesian information criterion (BIC).

\section{Results and analysis} \label{sec:result}
 We now proceed to compare the performance of our methods to the competing models we outlined in Section \ref{evaluation}. Table \ref{MDA} provides a summary of the forecasting ability of each model by presenting its MDA statistics for each cryptocurrency's return. The table includes four panels, each one presenting results for a different forecast horizon, with the rows focusing on the specific cryptocurrency. We begin by noting that both the LASSO-VAR are the most accurate in terms of MDA. Moreover, for the most stable variables only (such as USDT),homoscedastic model performance is comparable with the FGLS-VAR which on average performs better. Next, it appears that forecast performances increases with the forecast horizon.
 Furthermore, we do not see significant differences in performances among cryptocurrenceis belonging to the three tiers defined above.
 \begin{table}[H]
\caption{\scriptsize The table shows the Mean Directional Accuracy of model $i$ for variable $j$ computed as $MDA = \frac{1}{N}\sum^{\bar{t}}_{t=\underline{t}}sign(\mathrm{y_{{t+1},i,j}-y_{{t},i,j})}==sign(\mathrm{f_{{t+1},i,j}-f_{{t},i,j})}$. Where $N = 29$ are the directions for 30 days forecast. $y$ is the actual value and $f$ is the forecast. Bold values are the best results.}
\label{MDA}
\begin{adjustbox}{ max width=1\textwidth, center}
\begin{tabular}{clllllcllll}
\cline{1-5} \cline{7-11}
\multicolumn{1}{l}{Variable} & FGLS LASSO-VAR & OLS LASSO-VAR & LBVAR         & FAVAR         &  & \multicolumn{1}{l}{Variable} & FGLS LASSO-VAR & OLS LASSO-VAR & LBVAR         & FAVAR         \\ \cline{1-5} \cline{7-11} 
\multicolumn{1}{l}{}         & h=1            &               &               &               &  & \multicolumn{1}{l}{}         & h=2            &               &               &               \\ \cline{1-5} \cline{7-11} 
BTC-USD                      & \textbf{0.56}  & 0.48          & 0.46          & 0.46          &  & BTC-USD                      & \textbf{0.54}  & 0.50          & 0.50          & 0.43          \\
ETH-USD                      & \textbf{0.68}  & 0.40          & 0.50          & 0.46          &  & ETH-USD                      & \textbf{0.69}  & 0.50          & 0.61          & 0.54          \\
USDT-USD                     & 0.32           & \textbf{0.52} & 0.39          & 0.50          &  & USDT-USD                     & 0.38           & \textbf{0.54} & 0.46          & 0.54          \\
BNB-USD                      & 0.60           & \textbf{0.64} & 0.57          & 0.32          &  & BNB-USD                      & 0.62           & \textbf{0.65} & 0.50          & 0.32          \\
LTC-USD                      & 0.52           & 0.36          & \textbf{0.54} & 0.36          &  & LTC-USD                      & \textbf{0.54}  & 0.38          & 0.54          & 0.43          \\
ENJ-USD                      & 0.60           & \textbf{0.64} & 0.50          & 0.46          &  & ENJ-USD                      & \textbf{0.65}  & 0.62          & 0.46          & 0.39          \\
ZEN-USD                      & \textbf{0.56}  & 0.56          & 0.46          & 0.54          &  & ZEN-USD                      & 0.54           & 0.54          & 0.54          & \textbf{0.57} \\
NMC-USD                      & 0.48           & 0.52          & 0.61          & \textbf{0.64} &  & NMC-USD                      & 0.46           & 0.50          & 0.57          & \textbf{0.64} \\
PPC-USD                      & \textbf{0.64}  & 0.64          & 0.36          & 0.50          &  & PPC-USD                      & 0.62           & \textbf{0.65} & 0.39          & 0.50          \\
FTC-USD                      & 0.52           & 0.48          & 0.50          & \textbf{0.68} &  & FTC-USD                      & 0.54           & 0.54          & 0.46          & \textbf{0.68} \\ \cline{2-5} \cline{7-11} 
\multicolumn{1}{l}{}         & h=3            &               &               &               &  & \multicolumn{1}{l}{}         & h=4            &               &               &               \\ \cline{2-5} \cline{7-11} 
BTC-USD                      & \textbf{0.56}  & 0.44          & 0.50          & 0.43          &  & BTC-USD                      & \textbf{0.57}  & 0.43          & 0.50          & 0.46          \\
ETH-USD                      & \textbf{0.67}  & 0.52          & 0.61          & 0.54          &  & ETH-USD                      & \textbf{0.68}  & 0.57          & 0.61          & 0.57          \\
USDT-USD                     & 0.44           & \textbf{0.59} & 0.36          & 0.50          &  & USDT-USD                     & 0.50           & \textbf{0.57} & 0.36          & 0.50          \\
BNB-USD                      & \textbf{0.63}  & \textbf{0.63} & 0.50          & 0.32          &  & BNB-USD                      & \textbf{0.64}  & 0.61          & 0.50          & 0.32          \\
LTC-USD                      & 0.52           & 0.37          & \textbf{0.68} & 0.43          &  & LTC-USD                      & 0.54           & 0.36          & \textbf{0.64} & 0.46          \\
ENJ-USD                      & \textbf{0.67}  & 0.56          & 0.46          & 0.39          &  & ENJ-USD                      & \textbf{0.64}  & 0.57          & 0.50          & 0.39          \\
ZEN-USD                      & 0.52           & 0.52          & 0.46          & \textbf{0.54} &  & ZEN-USD                      & 0.46           & 0.50          & 0.54          & \textbf{0.54} \\
NMC-USD                      & 0.41           & 0.48          & 0.57          & \textbf{0.61} &  & NMC-USD                      & 0.39           & 0.50          & 0.57          & \textbf{0.68} \\
PPC-USD                      & 0.63           & \textbf{0.67} & 0.43          & 0.57          &  & PPC-USD                      & 0.61           & \textbf{0.64} & 0.43          & 0.57          \\
FTC-USD                      & 0.59           & 0.52          & 0.46          & \textbf{0.64} &  & FTC-USD                      & 0.61           & 0.57          & 0.54          & \textbf{0.68} \\ \hline
\end{tabular}
\end{adjustbox}
\end{table}
Table \ref{average} shows average results of each model considering also the heteroscedastic model without sentiment and search engine data (FGLS LASSO-VAR). We can see an average 10\% points improvement in directional accuracy of LASSO-VARs compared to benchmark models and a 37\% points increase compared to the model estimated without sentiment and Google trends.
\begin{table}[H]
\caption{\footnotesize Average MDA scores obtained from each model tested. Bold numbers represent the best results in a row. }
\label{average}
\begin{adjustbox}{ max width=1\textwidth, center}
\begin{tabular}{llllll}
\hline
        & FGLS LASSO-VAR  & OLS LASSO-VAR & \begin{tabular}[c]{@{}l@{}}FGLS LASSO-VAR \\ (without sentiment and gtrends)\end{tabular} & LBVAR  & FAVAR  \\ \hline
MDA h=1 & \textbf{0.5480} & 0.5240        & 0.3400                                                                                    & 0.4893 & 0.4929 \\
MDA h=2 & \textbf{0.5577} & 0.5423        & 0.3538                                                                                    & 0.5036 & 0.5036 \\
MDA h=3 & \textbf{0.5630} & 0.5296        & 0.3556                                                                                    & 0.5036 & 0.4964 \\
MDA h=4 & \textbf{0.5643} & 0.5321        & 0.3571                                                                                    & 0.5179 & 0.5179 \\ \hline
\end{tabular}
\end{adjustbox}
\end{table}
 Table \ref{RMSE} shows the forecast performances in terms of RMSE. In this case LBVAR gives us most precise results in absolute terms but they are not statistically different from those obtained with OLS and FGLS VARs in terms of EPA tests. This result may be linked to the possibility of this bayesian model to define a-priori a general as well as particular shrinkage intensity for each parameter ending up with more accurate estimates. On the contrary, both LASSO regularized models perform significantly better than FAVAR in terms of EPA tests.

\begin{table}[H]
\caption{\scriptsize The table shows the RMSE of model $i$ computed as $RMSE_{ij}=\sqrt{\sum^{\bar{t}-h}_{\tau=\underline{t}}e^2_{i,j,\tau+h}}$. where $e^2_{i,j,\tau+h}$ are the squared forecast error of variable $j$ at time $\tau$ and forecast horizon $h=1,\ldots,4$ generated by model $i$. $\bar{t}$ and $\underline{t}$ denote the start and the end of the out-of-sample period. All forecast are generated out-of-sample using recursive estimate of the model, with $\underline{t}=01/01/2022$ and $\bar{t}=01/31/2022$. Bold number represent the best results in a row. * significance at the 10\% level; ** significance at the 5\% level; *** significance at the 1\% level}
\label{RMSE}
\begin{adjustbox}{ max width=1\textwidth, center}
\begin{tabular}{clllllcllll}
\cline{1-5} \cline{7-11}
\multicolumn{1}{l}{Variable} & FGLS LASSO-VAR & OLS LASSO-VAR & LBVAR           & FAVAR  &  & \multicolumn{1}{l}{Variable} & FGLS LASSO-VAR & OLS LASSO-VAR & LBVAR           & FAVAR  \\ \cline{1-5} \cline{7-11} 
\multicolumn{1}{l}{}         & h=1            &               &                 &        &  & \multicolumn{1}{l}{}         & h=2            &               &                 &        \\ \cline{1-5} \cline{7-11} 
BTC-USD                      & 0.0585{***}         & 0.0527{***}         & \textbf{0.0310} & 0.1896 &  & BTC-USD                      & 0.0588{***}          & 0.0522{***}         & \textbf{0.0302} & 0.1604 \\
ETH-USD                      & 0.0660{***}          & 0.0646{***}        & \textbf{0.0484} & 0.2277 &  & ETH-USD                      & 0.0654{***}          & 0.0641{**}         & \textbf{0.0472} & 0.1869 \\
USDT-USD                     & 0.0055{***}          & 0.0041{***}         & \textbf{0.0003} & 0.1874 &  & USDT-USD                     & 0.0058{***}          & 0.0041{***}         & \textbf{0.0002} & 0.1925 \\
BNB-USD                      & 0.0703{***}          & 0.0629{***}         & \textbf{0.0501} & 0.1898 &  & BNB-USD                      & 0.0715{***}          & 0.0667{***}         & \textbf{0.0488} & 0.1682 \\
LTC-USD                      & 0.0853{***}          & 0.0720{***}         & \textbf{0.0454} & 0.2177 &  & LTC-USD                      & 0.0834{***}          & 0.0712{***}         & \textbf{0.0444} & 0.1560 \\
ENJ-USD                      & 0.1339{***}          & 0.1091{***}         & \textbf{0.0719} & 0.2731 &  & ENJ-USD                      & 0.1325{***}          & 0.1149{***}         & \textbf{0.0705} & 0.1944 \\
ZEN-USD                      & 0.0938{***}          & 0.0899{***}         & \textbf{0.0602} & 0.2800 &  & ZEN-USD                      & 0.0935{***}          & 0.0912{***}         & \textbf{0.0594} & 0.2045 \\
NMC-USD                      & 0.1342{***}          & 0.1309{***}         & \textbf{0.0634} & 0.2501 &  & NMC-USD                      & 0.1490         & 0.1456        & \textbf{0.0628} & 0.1838 \\
PPC-USD                      & 0.0691{***}          & 0.0675{***}         & \textbf{0.0412} & 0.1750 &  & PPC-USD                      & 0.0810{***}          & 0.0767{***}         & \textbf{0.0446} & 0.1626 \\
FTC-USD                      & 0.1275{*}          & 0.1186{***}         & \textbf{0.0676} & 0.1688 &  & FTC-USD                      & 0.1241         & 0.1169        & \textbf{0.0691} & 0.1527 \\ \cline{2-5} \cline{7-11} 
\multicolumn{1}{l}{}         & h=3            &               &                 &        &  & \multicolumn{1}{l}{}         & h=4            &               &                 &        \\ \cline{2-5} \cline{7-11} 
BTC-USD                      & 0.0425{***}          & 0.0499{***}         & \textbf{0.0296} & 0.1544 &  & BTC-USD                      & 0.0358{***}          & 0.0489{***}         & \textbf{0.0292} & 0.1463 \\
ETH-USD                      & 0.0544{**}          & 0.0555{**}         & \textbf{0.0463} & 0.1669 &  & ETH-USD                      & 0.0505{**}          & 0.0521{**}         & \textbf{0.0459} & 0.1484 \\
USDT-USD                     & 0.0040{***}          & 0.0041{***}         & \textbf{0.0002} & 0.2035 &  & USDT-USD                     & 0.0034{***}          & 0.0041{***}         & \textbf{0.0002} & 0.1935 \\
BNB-USD                      & 0.0654{***}          & 0.0644{***}         & \textbf{0.0482} & 0.1660 &  & BNB-USD                      & 0.0654{***}          & 0.0646{***}         & \textbf{0.0474} & 0.1567 \\
LTC-USD                      & 0.0630{***}          & 0.0677{***}         & \textbf{0.0436} & 0.1466 &  & LTC-USD                      & 0.0541{***}          & 0.0655{***}         & \textbf{0.0429} & 0.1389 \\
ENJ-USD                      & 0.1013{***}          & 0.1100{***}         & \textbf{0.0690} & 0.1777 &  & ENJ-USD                      & 0.0878{***}          & 0.1074{***}         & \textbf{0.0684} & 0.1650 \\
ZEN-USD                      & 0.0838{***}          & 0.0867{***}         & \textbf{0.0583} & 0.1900 &  & ZEN-USD                      & 0.0795{***}          & 0.0848{***}         & \textbf{0.0573} & 0.1773 \\
NMC-USD                      & 0.1072{*}          & 0.1121        & \textbf{0.0626} & 0.1562 &  & NMC-USD                      & 0.0939{*}          & 0.1042        & \textbf{0.0619} & 0.1458 \\
PPC-USD                      & 0.0806{**}          & 0.0761{**}         & \textbf{0.0464} & 0.1412 &  & PPC-USD                      & 0.0848{**}          & 0.0831{***}          & \textbf{0.0494} & 0.1396 \\
FTC-USD                      & 0.1029         & 0.1042        & \textbf{0.0677} & 0.1374 &  & FTC-USD                      & 0.0972         & 0.1007        & \textbf{0.0676} & 0.1225 \\ \hline
\end{tabular}
\end{adjustbox}
\end{table}

Finally, Table \ref{RMSE average} shows how sentiment variables are not contributing in achieving better results in terms of RMSE. This finding may be linked to the nature of sentiment indexes themselves. Being indexes, sentiments are not introducing into the model precise information regarding the out of sample actual value of the returns; they give, instead, information about the direction of our target variables. This is translated in better performance in terms of MDA as shown in Table \ref{average} but not in RMSE.

\begin{table}[H]
\caption{\footnotesize Average RMSE scores obtained from each model tested. Bold numbers represent the best results in a row. }
\label{RMSE average}
\begin{adjustbox}{ max width=1\textwidth, center}
\begin{tabular}{lllllll}
\hline
         & FGLS LASSO-VAR & OLS LASSO-VAR & \begin{tabular}[c]{@{}l@{}}FGLS LASSO-VAR\\ (without sentiment)\end{tabular} & \begin{tabular}[c]{@{}l@{}}FGLS LASSO-VAR\\ (without sentiment and gtrends)\end{tabular} & LBVAR           & FAVAR  \\ \hline
RMSE h=1 & 0.0844         & 0.0772        & 0.0818                                                                       & 0.0894                                                                                   & \textbf{0.0479} & 0.2159 \\
RMSE h=2 & 0.0865         & 0.0804        & 0.0849                                                                       & 0.0896                                                                                   & \textbf{0.0477} & 0.1762 \\
RMSE h=3 & 0.0705         & 0.0731        & 0.0733                                                                       & 0.0785                                                                                   & \textbf{0.0472} & 0.1640 \\
RMSE h=4 & 0.0652         & 0.0715        & 0.0698                                                                       & 0.0739                                                                                   & \textbf{0.0470} & 0.1534 \\ \hline
\end{tabular}
\end{adjustbox}
\end{table}

\begin{figure}[H]
    \centering
    \includegraphics[width=16cm,height=10cm]{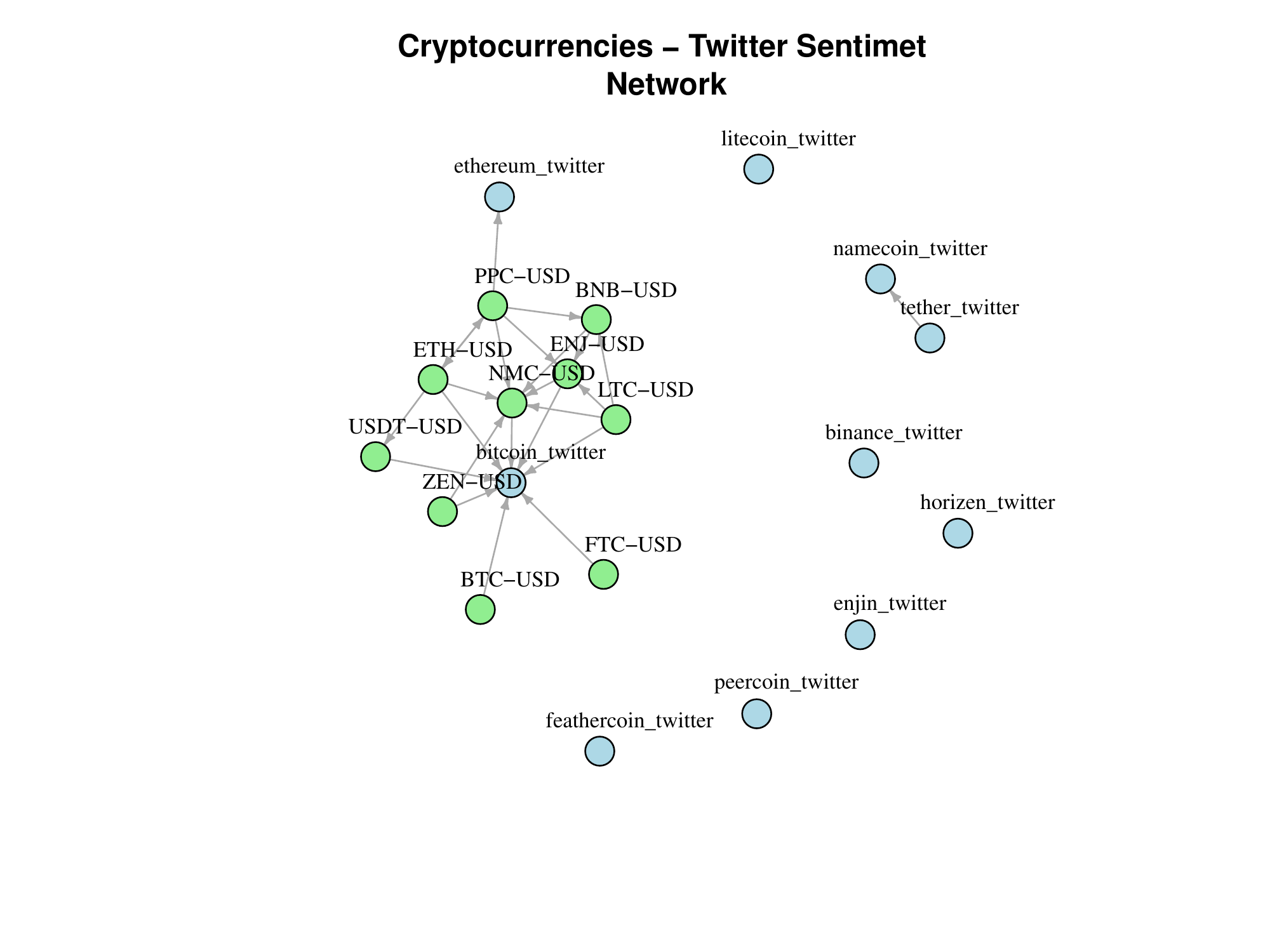}
    \vspace{-2cm}
    \caption{\footnotesize The figure shows a network between Twitter sentiment indexes and returns.The direction of the arrows represents the direction of the Granger causality}
    \label{twit_net}
\end{figure}

\begin{figure}[H]
    \centering
    \includegraphics[width=16cm,height=10cm]{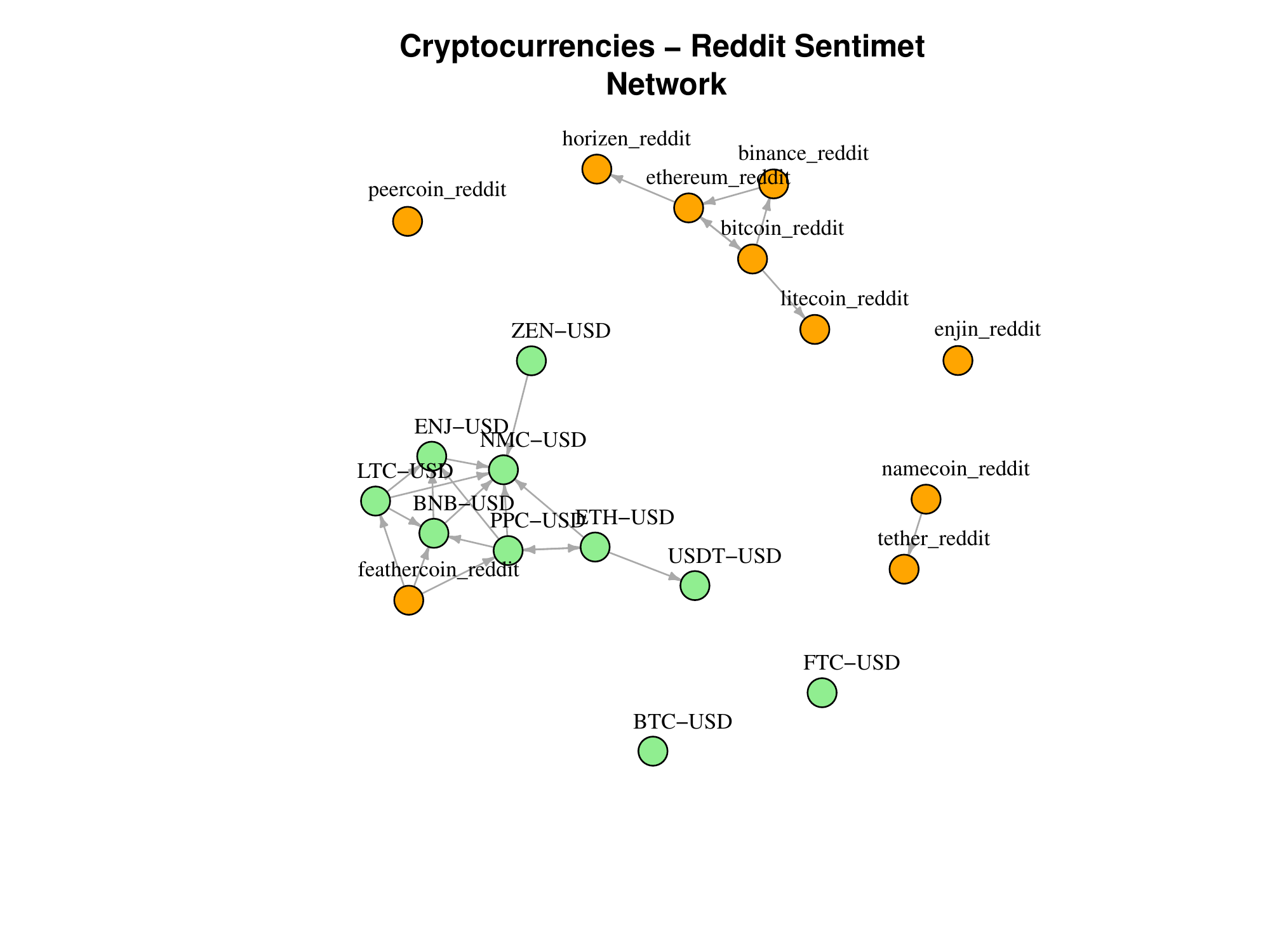}
    \vspace{-2cm}
    \caption{\footnotesize The figure shows a network between Reddit sentiment indexes and returns.The direction of the arrows represents the direction of the Granger causality}
    \label{red_net}
\end{figure}

\begin{figure}[H]
    \centering
    \includegraphics[width=16cm,height=10cm]{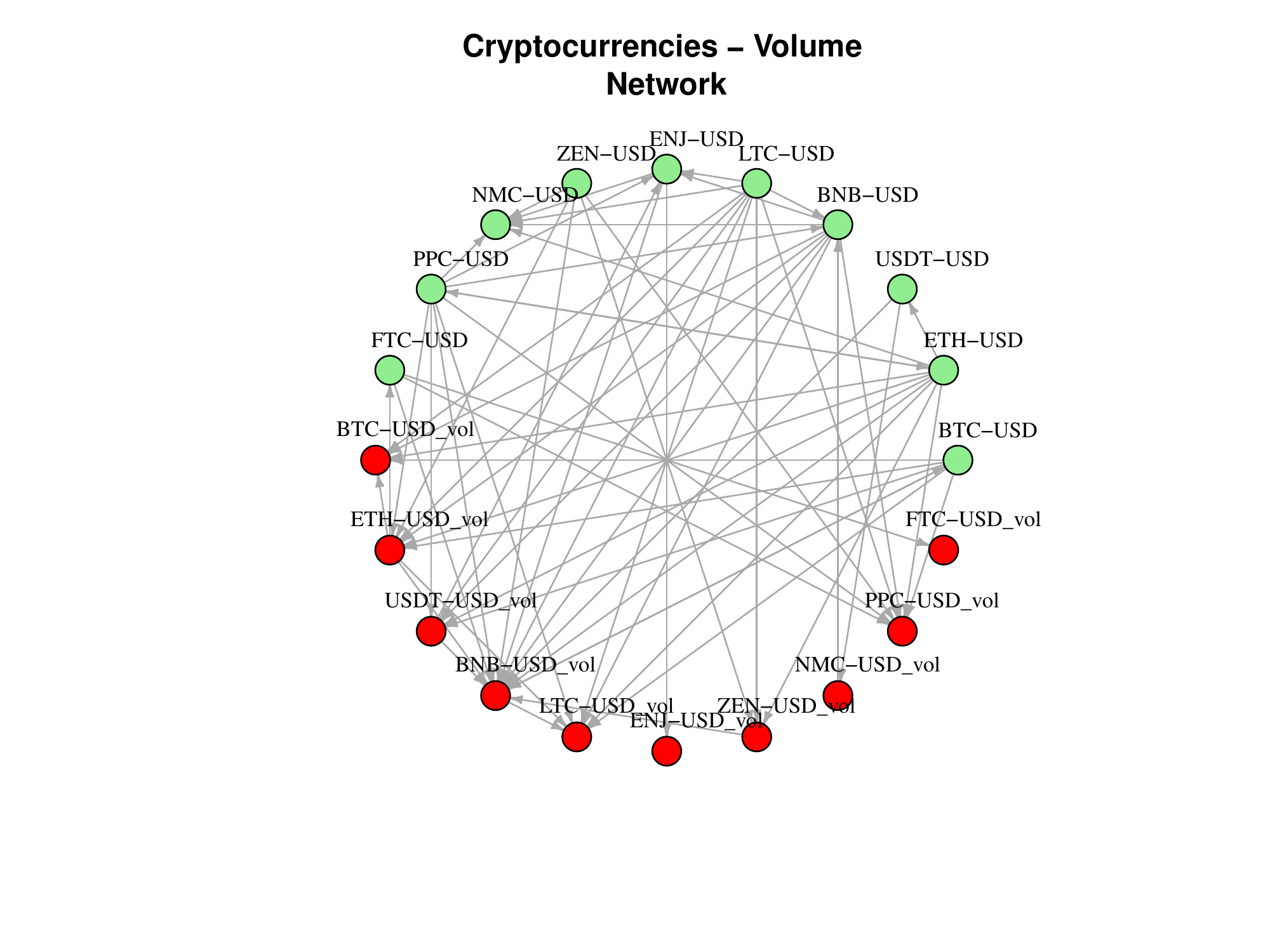}
    \vspace{-2cm}
    \caption{\footnotesize The figure shows a network between Volume and returns.The direction of the arrows represents the direction of the Granger causality}
    \label{vol_net}
\end{figure}

\clearpage
Concerning Granger causality, we report in Figures \ref{twit_net}, \ref{red_net}, and \ref{vol_net} networks of returns and sentiment and volume variables. Arrows are depicted when the null-hypothesis is rejected with the  $\textit{p-values}$ of LM test $<0.01$. The direction of the arrows represents the one of the Granger causality. We report the $\textit{p-values}$ for each combination tested in the Appendix Table \ref{GCtable}\footnote{For simplicity we did not report all the combinations when the Granger causality is tested from Google trends. However, $\textit{p-values}$ reported are computed considering also search engine data as ``other variables" }.
From Figure \ref{twit_net}, we can conclude that there is no Granger causality between Twitter sentiments and cryptocurrencies returns. On the contrary, returns seem to cause each other, and most of them are causing the Bitcoin sentiment extracted from Twitter. From this finding, it is reasonable to interpret Bitcoin Twitter sentiment as a general index for all cryptocurrencies market. It is also interesting to note that the biggest and the smallest cryptocurrency in market capitalization are outside the cluster created by all the other currencies. 

Same results are obtained focusing on the relation between Reddit sentiments and returns. Figure \ref{red_net} shows, in fact, that sentiments obtained from Reddit are not Granger causing their respective cryptocurrency. Also, in this case, Bitcoin, Feathercoin, and the stablecoin Tether are outside the cluster created by the other currencies. It is interesting to focus on the cluster formed by several sentiments around the Bitcoin one. This finding may describe that the euphoria in specific currencies Sub-reddit is linked to the users' feelings expressed in the Bitcoin Sub-reddit, which pushes the others.
Figure \ref{vol_net} shows the Granger causality network between returns and volume. 
In this case, we can see that returns are Granger causing currencies volume. These results suggest that movements in cryptocurrency returns may influence the market of the others, primarily via demand shocks. Finally no causality is found from Google trends and cryptocurrencies return.

\section{Conclusion} \label{sec:conclusion} 
 We constructed a novel dataset combining cryptocurrencies returns, social media sentiment indexes, Google trends, and volume. We wanted to analyze the forecast performance of regularized high-dimensional VAR models exploiting the novel dataset we created.
 In a thirty days forecast, we obtain a clear improvement in MDA by comparing our results with the most popular high-dimensional VAR models. Moreover, we observe comparable and precise performance in terms of RMSE considering LBVAR as benchmark. In light of these results, we can conclude that high-dimensional VAR models are good tools to model highly volatile financial time series such as cryptocurrency returns. Furthermore, the analysis of forecast performance metrics suggests a significant contribution of sentiment and attention measures in achieving better results in terms of directional accuracy. This finding can be decisive for portfolio selection strategies or, in general, for investment purposes.
 The contribution of sentiment and search engine variables is not translated into Granger causality. In most cases, we cannot reject the null hypothesis of (non-)Granger causality. However, we find an interesting connection between volume and cryptocurrency returns.

\spacing{1}
\clearpage
\phantomsection
\bibliography{ref}
\spacing{1.2}
\clearpage
\appendix
\section{Appendix}
\label{Appendix}
\subsection{Google trends}
\begin{table}[H]
\caption{\footnotesize Google Trends collected from January 2018 to January 2022.} 
\label{gtrends}
\centering
\begin{tabular}{ll}
\hline
\multicolumn{2}{c}{\textbf{Google Trends}}            \\ \hline
\textit{binance coin}   & \textit{zen coin}           \\
\textit{crypto crash}   & \textit{coinbase}           \\
\textit{miner}          & \textit{BTC}                \\
\textit{USD coin}       & \textit{hard fork}          \\
\textit{cardano}        & \textit{stablecoin}         \\
\textit{cryptocurrency} & \textit{cold storage}       \\
\textit{minted}         & \textit{coin}               \\
\textit{solana}         & \textit{hash}               \\
\textit{feathercoin}    & \textit{tether}             \\
\textit{cryptography}   & \textit{litecoin}           \\
\textit{dogecoin}       & \textit{enjin coin}         \\
\textit{digital gold}   & \textit{hashing}            \\
\textit{Bitcoin}        & \textit{token}              \\
\textit{NFT}            & \textit{crypto}             \\
\textit{ripple}         & \textit{ICO}                \\
\textit{private key}    & \textit{coin market}        \\
\textit{peercoin}       & \textit{soft fork}          \\
\textit{namecoin}       & \textit{block producer}     \\
\textit{satoshi}        & \textit{distributed ledger} \\
\textit{hot wallet}     & \textit{digital fiat}       \\
\textit{blockchain}     & \textit{consensus}          \\
\textit{ethereum}       & \textit{\textbf{}}          \\ \hline
\end{tabular}
\end{table}

\subsection{Granger causality}
\begin{table}[]
\caption{$\textit{p-values}$ of Post-double-selection Granger causality LM test from cryptocurrencies returns to other variables }
\label{GCtable}
\begin{adjustbox}{ max width=0.8\textwidth, center}
\begin{tabular}{lllllllllll}
\hline
GCto↓ \quad                GCfrom→ & BTC-USD         & ETH-USD         & USDT-USD        & BNB-USD         & LTC-USD         & ENJ-USD         & ZEN-USD         & NMC-USD         & PPC-USD         & FTC-USD         \\ \hline
BTC-USD                     & NA              & 0.0160          & 0.5806          & 0.0766          & 0.0480          & 0.8495          & 0.1692          & 0.2867          & 0.0509          & 0.3835          \\
ETH-USD                     & 0.2424          & NA              & 0.1301          & 0.1656          & 0.6113          & 0.7493          & 0.2147          & 0.0308          & \textbf{0.0095} & 0.6677          \\
USDT-USD                    & 0.0304          & \textbf{0.0082} & NA              & 0.0849          & 0.0607          & 0.0530          & 0.1661          & 0.1990          & 0.0586          & 0.0592          \\
BNB-USD                     & 0.4005          & 0.0990          & 0.2285          & NA              & \textbf{0.0099} & 0.0953          & 0.1949          & 0.1746          & \textbf{0.0018} & 0.0100          \\
LTC-USD                     & 0.1818          & 0.0343          & 0.1641          & 0.1758          & NA              & 0.3354          & 0.0577          & 0.1425          & 0.0348          & 0.0694          \\
ENJ-USD                     & 0.0634          & 0.0180          & 0.8120          & \textbf{0.0017} & \textbf{0.0005} & NA              & 0.5604          & 0.0110          & \textbf{0.0000} & 0.2315          \\
ZEN-USD                     & 0.0762          & 0.0439          & 0.1940          & 0.5320          & 0.2329          & 0.0437          & NA              & 0.2635          & 0.0713          & 0.8826          \\
NMC-USD                     & 0.0251          & \textbf{0.0002} & 0.0703          & \textbf{0.0000} & \textbf{0.0009} & \textbf{0.0011} & \textbf{0.0007} & NA              & \textbf{0.0001} & 0.1259          \\
PPC-USD                     & 0.1485          & \textbf{0.0054} & 0.5124          & 0.0277          & 0.0392          & 0.4461          & 0.1795          & 0.9247          & NA              & 0.0804          \\
FTC-USD                     & 0.1446          & 0.0309          & 0.8128          & 0.0260          & 0.0107          & 0.7316          & 0.0719          & 0.9981          & 0.0158          & NA              \\
binance\_twitter            & 0.3954          & 0.2646          & 0.6356          & 0.6085          & 0.4001          & 0.3064          & 0.2354          & 0.9448          & 0.4448          & 0.6705          \\
bitcoin\_twitter            & \textbf{0.0037} & \textbf{0.0004} & \textbf{0.0005} & 0.0294          & \textbf{0.0085} & \textbf{0.0002} & \textbf{0.0004} & \textbf{0.0051} & 0.0489          & \textbf{0.0097} \\
enjin\_twitter              & 0.8122          & 0.7953          & 0.1428          & 0.3382          & 0.6695          & 0.8774          & 0.2228          & 0.7266          & 0.9771          & 0.4733          \\
ethereum\_twitter           & 0.5745          & 0.5471          & 0.4612          & 0.1124          & 0.5614          & 0.6184          & 0.4833          & 0.2719          & \textbf{0.0030} & 0.9763          \\
feathercoin\_twitter        & 0.1760          & 0.4113          & 0.2249          & 0.8650          & 0.2708          & 0.5450          & 0.2588          & 0.6098          & 0.5435          & 0.3414          \\
horizen\_twitter            & 0.3166          & 0.7868          & 0.1714          & 0.1453          & 0.2374          & 0.3077          & 0.4665          & 0.6470          & 0.3518          & 0.5166          \\
litecoin\_twitter           & 0.4395          & 0.4252          & 0.4043          & 0.9540          & 0.3795          & 0.1323          & 0.1898          & 0.4090          & 0.6604          & 0.1760          \\
namecoin\_twitter           & 0.8102          & 0.5725          & 0.6608          & 0.5611          & 0.8145          & 0.1292          & 0.0801          & 0.0839          & 0.8672          & 0.1424          \\
peercoin\_twitter           & 0.4027          & 0.5644          & 0.5816          & 0.2592          & 0.7573          & 0.8720          & 0.2728          & 0.8352          & 0.4572          & 0.6181          \\
tether\_twitter             & 0.0963          & 0.4915          & 0.0152          & 0.5721          & 0.5265          & 0.0875          & 0.7624          & 0.5208          & 0.0772          & 0.3862          \\
binance\_reddit             & 0.6541          & 0.6082          & 0.6289          & 0.2662          & 0.9793          & 0.5826          & 0.9291          & 0.8083          & 0.5520          & 0.6791          \\
bitcoin\_reddit             & 0.0672          & 0.0673          & 0.8790          & 0.1642          & 0.0385          & 0.2775          & 0.0991          & 0.1266          & 0.0427          & 0.4702          \\
enjin\_reddit               & 0.6851          & 0.4729          & 0.4735          & 0.5772          & 0.8391          & 0.9730          & 0.2725          & 0.0446          & 0.6572          & 0.2913          \\
ethereum\_reddit            & 0.1089          & 0.2272          & 0.2207          & 0.7603          & 0.7171          & 0.7370          & 0.4804          & 0.3783          & 0.5760          & 0.3082          \\
feathercoin\_reddit         & 0.0561          & 0.1935          & 0.9954          & 0.0507          & 0.0470          & 0.4072          & 0.0116          & 0.2093          & 0.0256          & 0.2115          \\
horizen\_reddit             & 0.5628          & 0.7018          & 0.5875          & 0.3949          & 0.3620          & 0.7171          & 0.9944          & 0.6513          & 0.6111          & 0.9837          \\
litecoin\_reddit            & 0.7097          & 0.4810          & 0.3864          & 0.8410          & 0.5646          & 0.5027          & 0.5521          & 0.9301          & 0.9295          & 0.8797          \\
namecoin\_reddit            & 0.3928          & 0.0413          & 0.1420          & 0.2572          & 0.1178          & 0.3858          & 0.5974          & 0.5896          & 0.0552          & 0.0687          \\
peercoin\_reddit            & 0.6658          & 0.7312          & 0.5510          & 0.6567          & 0.4873          & 0.6190          & 0.6782          & 0.1117          & 0.6449          & 0.7178          \\
tether\_reddit              & 0.3356          & 0.3148          & 0.4185          & 0.5696          & 0.4660          & 0.9597          & 0.2934          & 0.1441          & 0.0401          & 0.9379          \\
binance coin                & \textbf{0.0000} & \textbf{0.0000} & 0.6609          & \textbf{0.0000} & \textbf{0.0000} & 0.0834          & \textbf{0.0048} & 0.2576          & 0.1140          & 0.2205          \\
crypto crash                & \textbf{0.0000} & \textbf{0.0000} & 0.9778          & \textbf{0.0000} & \textbf{0.0000} & \textbf{0.0000} & \textbf{0.0000} & 0.0462          & \textbf{0.0000} & 0.0191          \\
miner                       & \textbf{0.0041} & 0.0322          & 0.4051          & 0.0505          & \textbf{0.0067} & 0.2852          & 0.1471          & 0.7523          & 0.6174          & 0.7578          \\
USD coin                    & \textbf{0.0034} & \textbf{0.0046} & 0.4258          & \textbf{0.0045} & \textbf{0.0001} & 0.7856          & 0.0409          & 0.4670          & 0.3972          & \textbf{0.0062} \\
cardano                     & 0.1536          & 0.0814          & 0.4645          & \textbf{0.0067} & 0.1593          & 0.9221          & 0.5557          & 0.1927          & 0.0998          & 0.2494          \\
cryptocurrency              & \textbf{0.0006} & \textbf{0.0001} & 0.3861          & \textbf{0.0000} & \textbf{0.0001} & \textbf{0.0009} & \textbf{0.0065} & 0.0107          & \textbf{0.0018} & 0.1033          \\
minted                      & 0.4002          & 0.2837          & 0.8245          & 0.1180          & 0.4083          & 0.0446          & 0.5992          & 0.7046          & 0.9410          & 0.4048          \\
solana                      & 0.7416          & 0.5055          & 0.3076          & 0.0797          & 0.3734          & 0.2871          & 0.3570          & 0.7301          & 0.9052          & 0.5513          \\
feathercoin                 & 0.0862          & 0.0739          & 0.1444          & 0.6347          & \textbf{0.0005} & 0.1046          & 0.5542          & \textbf{0.0006} & 0.7720          & 0.0682          \\
cryptography                & 0.1061          & 0.0340          & 0.4799          & 0.1491          & 0.1080          & 0.4966          & 0.6945          & 0.7332          & 0.8046          & 0.2341          \\
dogecoin                    & 0.3726          & 0.1008          & 0.9989          & 0.4082          & 0.3151          & 0.5171          & 0.5089          & 0.8553          & 0.2065          & 0.2517          \\
digital gold                & 0.3546          & 0.2963          & 0.3691          & 0.7348          & 0.4403          & 0.2610          & 0.8192          & 0.9313          & 0.2141          & 0.2325          \\
Bitcoin                     & 0.0159          & \textbf{0.0001} & 0.4836          & \textbf{0.0000} & \textbf{0.0063} & \textbf{0.0006} & \textbf{0.0075} & 0.1809          & 0.0182          & 0.0678          \\
NFT                         & 0.1294          & 0.0845          & 1.0000          & 0.0804          & 0.1433          & 0.0188          & 0.0391          & 0.8159          & 0.7121          & 0.1280          \\
ripple                      & 0.1088          & 0.1370          & 0.1399          & \textbf{0.0069} & 0.0273          & 0.3337          & 0.6380          & 0.1514          & 0.6041          & 0.9649          \\
private key                 & 0.2985          & 0.1430          & 0.5925          & 0.5632          & 0.4996          & 0.6208          & 0.3199          & 0.3138          & 0.3721          & 0.3507          \\
peercoin                    & 0.7811          & 0.2091          & 0.6699          & 0.9692          & 0.3320          & 0.8001          & 0.9222          & 0.5993          & \textbf{0.0050} & 0.5869          \\
namecoin                    & 0.1008          & 0.0260          & 0.6049          & 0.1510          & 0.1160          & 0.1320          & 0.2546          & \textbf{0.0081} & 0.8135          & 0.0356          \\
satoshi                     & \textbf{0.0005} & 0.0130          & 0.0430          & 0.6887          & \textbf{0.0051} & 0.8851          & 0.3407          & 0.3459          & 0.2010          & \textbf{0.0012} \\
hot wallet                  & 0.0568          & 0.0244          & 0.1037          & 0.1298          & 0.0174          & 0.2064          & 0.8107          & 0.6797          & 0.1160          & 0.1919          \\
blockchain                  & \textbf{0.0000} & \textbf{0.0007} & 0.8780          & 0.1187          & \textbf{0.0000} & 0.4025          & 0.1450          & 0.1267          & 0.0590          & 0.3071          \\
ethereum                    & \textbf{0.0000} & \textbf{0.0000} & 0.8132          & 0.0235          & \textbf{0.0002} & 0.4269          & 0.1647          & 0.4441          & \textbf{0.0007} & 0.0779          \\
zen coin                    & 0.0903          & 0.3427          & 0.2019          & 0.3071          & 0.0986          & 0.0698          & 0.8930          & 0.7814          & 0.6367          & 0.8439          \\
coinbase                    & \textbf{0.0011} & 0.0294          & 0.7721          & 0.1764          & \textbf{0.0020} & 0.1930          & 0.4510          & 0.5053          & 0.2014          & 0.4402          \\
BTC                         & 0.1413          & \textbf{0.0050} & 0.5335          & 0.0199          & 0.1986          & \textbf{0.0085} & 0.2251          & 0.3561          & 0.3073          & 0.2519          \\
hard fork                   & 0.8211          & 0.9871          & 0.3466          & 0.6733          & 0.9744          & 0.9280          & 0.6540          & 0.4868          & 0.2830          & 0.7385          \\
stablecoin                  & 0.6050          & 0.1118          & 0.2014          & 0.3165          & 0.1344          & 0.8491          & 0.6773          & 0.2453          & 0.3065          & 0.7987          \\
cold storage                & 0.4250          & 0.7975          & 0.6901          & 0.1378          & 0.6385          & 0.6781          & 0.3314          & 0.0878          & 0.8518          & 0.1559          \\
coin                        & \textbf{0.0037} & 0.0118          & 0.3139          & 0.5546          & \textbf{0.0024} & 0.1975          & 0.0345          & 0.6244          & 0.9705          & 0.9465          \\
hash                        & 0.7259          & 0.5989          & 0.0102          & 0.4963          & 0.4929          & 0.9033          & 0.9252          & 0.9961          & 0.8192          & 0.0555          \\
tether                      & 0.6599          & 0.8826          & 0.7482          & 0.3878          & 0.7186          & 0.3158          & 0.8085          & 0.0353          & 0.4464          & 0.6137          \\
litecoin                    & \textbf{0.0072} & 0.2772          & 0.4344          & 0.1784          & \textbf{0.0010} & 0.5565          & 0.6718          & 0.1310          & 0.0648          & 0.1283          \\
enjin coin                  & \textbf{0.0007} & 0.0464          & 0.9766          & \textbf{0.0021} & \textbf{0.0012} & \textbf{0.0000} & 0.0711          & 0.1025          & 0.2130          & 0.8599          \\
hashing                     & 0.3379          & 0.3305          & 0.2290          & 0.3685          & 0.4969          & 0.5957          & 0.6382          & 0.0339          & 0.6142          & 0.3185          \\
token                       & 0.0498          & 0.0795          & 0.5437          & 0.1770          & 0.2085          & 0.6911          & 0.7638          & 0.1078          & 0.0178          & 0.0712          \\
crypto                      & \textbf{0.0000} & \textbf{0.0000} & 0.8925          & \textbf{0.0000} & \textbf{0.0000} & 0.0127          & \textbf{0.0013} & \textbf{0.0004} & \textbf{0.0002} & \textbf{0.0004} \\
ICO                         & 0.0145          & 0.0365          & 0.9128          & 0.2933          & 0.0713          & 0.0469          & 0.0234          & 0.4183          & 0.2966          & 0.0582          \\
coin market                 & \textbf{0.0026} & 0.0371          & 0.1630          & 0.1688          & \textbf{0.0008} & 0.2388          & 0.4455          & 0.3456          & 0.2171          & 0.8817          \\
soft fork                   & 0.3565          & 0.1827          & 0.8709          & 0.0512          & 0.0582          & 0.0308          & 0.4759          & 0.4489          & 0.5009          & 0.9645          \\
block producer              & 0.8401          & 0.9728          & 0.9367          & 0.5129          & 0.9480          & 0.8211          & 0.5828          & 0.3741          & 0.8435          & 0.4542          \\
distributed ledger          & 0.6765          & 0.9060          & 0.2695          & 0.9712          & 0.7463          & 0.3372          & 0.1931          & 0.7627          & 0.0919          & 0.3213          \\
digital fiat                & 0.7310          & 0.3724          & 0.0663          & 0.7566          & 0.7494          & 0.2539          & 0.5716          & 0.4768          & 0.9983          & 0.9122          \\
consensus                   & 0.9003          & 0.9690          & 0.9322          & 0.8829          & 0.8762          & 0.9336          & 0.7570          & 0.1679          & 0.8630          & 0.4639          \\
BTC-USD\_vol                & \textbf{0.0017} & \textbf{0.0005} & 0.2268          & \textbf{0.0012} & \textbf{0.0074} & 0.5861          & 0.2199          & 0.1915          & 0.0160          & 0.1666          \\
ETH-USD\_vol                & \textbf{0.0003} & \textbf{0.0000} & 0.1126          & \textbf{0.0071} & \textbf{0.0001} & 0.1695          & \textbf{0.0098} & 0.1317          & \textbf{0.0089} & 0.0692          \\
USDT-USD\_vol               & \textbf{0.0001} & \textbf{0.0000} & 0.0787          & \textbf{0.0051} & \textbf{0.0000} & 0.3197          & 0.0376          & 0.0227          & \textbf{0.0008} & 0.1977          \\
BNB-USD\_vol                & \textbf{0.0000} & \textbf{0.0000} & \textbf{0.0001} & \textbf{0.0000} & \textbf{0.0000} & \textbf{0.0000} & \textbf{0.0053} & 0.0854          & \textbf{0.0002} & \textbf{0.0000} \\
LTC-USD\_vol                & \textbf{0.0002} & \textbf{0.0003} & 0.1229          & \textbf{0.0037} & \textbf{0.0000} & 0.5721          & 0.0861          & 0.1896          & \textbf{0.0008} & 0.0123          \\
ENJ-USD\_vol                & 0.2829          & 0.2564          & 0.0412          & 0.0309          & 0.0304          & \textbf{0.0000} & 0.6242          & 0.0387          & 0.1828          & 0.4097          \\
ZEN-USD\_vol                & 0.0126          & \textbf{0.0008} & 0.3450          & 0.0705          & \textbf{0.0017} & 0.0130          & \textbf{0.0000} & 0.0137          & 0.0682          & 0.3457          \\
NMC-USD\_vol                & 0.0140          & 0.0242          & \textbf{0.0000} & 0.0132          & 0.0298          & 0.3640          & 0.0602          & 0.0255          & 0.2465          & 0.1979          \\
PPC-USD\_vol                & \textbf{0.0029} & \textbf{0.0001} & 0.0124          & \textbf{0.0013} & \textbf{0.0000} & 0.0176          & \textbf{0.0001} & 0.1828          & \textbf{0.0000} & \textbf{0.0037} \\
FTC-USD\_vol                & 0.0693          & 0.3107          & 0.0925          & 0.2897          & 0.0504          & 0.6788          & 0.0187          & 0.2529          & 0.4108          & \textbf{0.0006} \\ \hline
\end{tabular}
\end{adjustbox}
\end{table}
\begin{table}[]
\caption{$\textit{p-values}$ of sample corrected $\textit{F}-statistic$ of Granger (non-)causality test from Twitter sentiments to other variables }
\label{GCtable}
\begin{adjustbox}{ max width=1\textwidth, center}
\begin{tabular}{lllllllllll}
\hline
Gcto                 GCfrom & binance\_twitter & bitcoin\_twitter & enjin\_twitter  & ethereum\_twitter & feathercoin\_twitter & horizen\_twitter & litecoin\_twitter & namecoin\_twitter & peercoin\_twitter & tether\_twitter \\ \hline
BTC-USD                     & 0.2836           & 0.6193           & 0.4217          & 0.1768            & 0.7939               & 0.5917           & 0.5067            & 0.9076            & 0.1186            & 0.7629          \\
ETH-USD                     & 0.0445           & 0.0526           & 0.6375          & 0.0155            & 0.7203               & 0.7424           & 0.1068            & 0.1708            & 0.3052            & 0.6660          \\
USDT-USD                    & 0.9294           & 0.1183           & 0.0616          & 0.5228            & 0.0302               & 0.2846           & 0.4805            & 0.3773            & 0.6936            & 0.0230          \\
BNB-USD                     & 0.4527           & 0.1739           & 0.2443          & 0.1532            & 0.7731               & 0.7727           & 0.0739            & 0.0348            & 0.3130            & 0.6413          \\
LTC-USD                     & 0.1515           & 0.5947           & 0.5626          & 0.0249            & 0.5487               & 0.1943           & 0.1889            & 0.6979            & 0.1782            & 0.2934          \\
ENJ-USD                     & 0.5534           & 0.4781           & 0.3863          & 0.7358            & 0.9879               & 0.3070           & 0.0189            & 0.1175            & 0.3266            & 0.9704          \\
ZEN-USD                     & 0.7816           & 0.6695           & 0.8269          & 0.0797            & 0.9265               & 0.2341           & 0.4250            & 0.0847            & 0.2048            & 0.6214          \\
NMC-USD                     & 0.2995           & 0.6881           & 0.9317          & 0.4905            & 0.3274               & 0.0705           & 0.1071            & 0.8214            & 0.4319            & 0.7449          \\
PPC-USD                     & 0.6282           & 0.9610           & 0.3731          & 0.0633            & 0.2392               & 0.1032           & 0.8967            & 0.9697            & 0.1709            & 0.6841          \\
FTC-USD                     & 0.2389           & 0.3845           & 0.2214          & 0.0325            & 0.8066               & 0.5975           & 0.3111            & 0.1334            & 0.0210            & 0.6129          \\
binance\_twitter            & NA               & 0.0193           & 0.9603          & 0.0224            & 0.5917               & 0.7926           & 0.6675            & 0.9330            & 0.8544            & 0.2448          \\
bitcoin\_twitter            & 0.6348           & NA               & 0.7212          & 0.0927            & 0.4338               & 0.9155           & 0.8053            & 0.4358            & 0.2363            & 0.5879          \\
enjin\_twitter              & 0.2779           & 0.5965           & NA              & 0.9036            & 0.0439               & 0.1869           & 0.8787            & 0.3202            & 0.4747            & 0.1904          \\
ethereum\_twitter           & 0.4339           & 0.1282           & 0.0640          & NA                & 0.8284               & 0.9192           & 0.2065            & 0.0317            & 0.7232            & 0.1088          \\
feathercoin\_twitter        & 0.1484           & 0.5466           & 0.4829          & 0.9142            & NA                   & 0.9193           & 0.2521            & 0.0816            & 0.8440            & 0.0376          \\
horizen\_twitter            & 0.3591           & 0.3107           & 0.0328          & 0.4325            & 0.7878               & NA               & 0.6530            & 0.6684            & 0.2187            & 0.3314          \\
litecoin\_twitter           & 0.3215           & 0.3492           & 0.6308          & 0.8966            & 0.7400               & 0.8926           & NA                & 0.4861            & 0.6718            & 0.3686          \\
namecoin\_twitter           & 0.0851           & 0.5768           & 0.8819          & 0.2969            & 0.8555               & 0.2732           & 0.0134            & NA                & 0.0658            & \textbf{0.0089} \\
peercoin\_twitter           & 0.0561           & 0.6776           & 0.7402          & 0.2010            & 0.9291               & 0.4143           & 0.1932            & 0.8459            & NA                & 0.8278          \\
tether\_twitter             & 0.0958           & 0.0698           & 0.5065          & 0.2755            & 0.0764               & 0.9713           & 0.1445            & 0.5973            & 0.1683            & NA              \\
binance\_reddit             & 0.1073           & 0.2070           & 0.1834          & 0.9775            & 0.3258               & 0.9364           & 0.9399            & 0.8363            & 0.1013            & 0.8588          \\
bitcoin\_reddit             & 0.5128           & 0.3099           & 0.6017          & 0.7855            & \textbf{0.0081}      & 0.7087           & 0.2620            & 0.4051            & 0.4614            & 0.1572          \\
enjin\_reddit               & 0.1477           & 0.3237           & 0.7215          & 0.3622            & 0.2115               & 0.3067           & 0.8910            & 0.3101            & 0.2733            & 0.9930          \\
ethereum\_reddit            & 0.5897           & 0.6249           & 0.1164          & 0.4412            & 0.6651               & 0.8423           & 0.6207            & 0.3736            & 0.3048            & 0.1188          \\
feathercoin\_reddit         & 0.1224           & 0.5682           & 0.4413          & 0.7241            & 0.5178               & 0.3215           & 0.5211            & 0.0623            & 0.5782            & 0.3215          \\
horizen\_reddit             & 0.3031           & 0.2193           & 0.9052          & 0.4061            & 0.1514               & 0.6784           & 0.2436            & 0.4387            & 0.6865            & 0.7815          \\
litecoin\_reddit            & 0.9259           & 0.3443           & 0.0409          & 0.9613            & 0.4127               & 0.9871           & 0.0394            & \textbf{0.0051}   & 0.5193            & 0.9694          \\
namecoin\_reddit            & 0.2243           & 0.4472           & 0.1992          & 0.3831            & 0.9794               & 0.4073           & 0.1210            & 0.9300            & 0.1832            & 0.4928          \\
peercoin\_reddit            & 0.3292           & 0.3009           & 0.8619          & 0.8040            & 0.6110               & 0.1571           & 0.3354            & 0.3678            & 0.1571            & 0.7751          \\
tether\_reddit              & 0.5138           & 0.2140           & 0.5892          & 0.3380            & 0.5673               & 0.6192           & 0.2542            & 0.5848            & 0.2780            & 0.4751          \\
binance coin                & 0.8862           & 0.1407           & 0.5714          & 0.9191            & 0.9673               & 0.8414           & 0.2558            & 0.6459            & 0.3450            & 0.6336          \\
crypto crash                & 0.3984           & \textbf{0.0002}  & 0.5467          & \textbf{0.0029}   & 0.8798               & 0.8268           & 0.2549            & 0.7603            & \textbf{0.0052}   & 0.7470          \\
miner                       & 0.0502           & 0.1712           & 0.4746          & 0.5992            & 0.0925               & 0.5411           & 0.9161            & 0.9501            & 0.4062            & 0.1135          \\
USD coin                    & 0.8942           & \textbf{0.0000}  & 0.8090          & 0.6022            & 0.6801               & 0.6448           & \textbf{0.0093}   & 0.6302            & 0.8233            & 0.3079          \\
cardano                     & 0.7775           & 0.0588           & 0.3634          & \textbf{0.0061}   & 0.6833               & 0.6337           & \textbf{0.0019}   & 0.2495            & 0.1241            & 0.5301          \\
cryptocurrency              & 0.4750           & 0.1088           & 0.2338          & 0.6727            & 0.7848               & 0.9204           & 0.1712            & 0.8256            & 0.0568            & 0.1901          \\
minted                      & 0.4680           & 0.0388           & 0.1532          & 0.7794            & 0.3114               & 0.5892           & 0.4885            & 0.2293            & 0.5396            & 0.2287          \\
solana                      & 0.2837           & 0.3680           & 0.1212          & 0.1947            & 0.7989               & 0.8472           & 0.5140            & 0.5022            & 0.6212            & 0.4513          \\
feathercoin                 & 0.0506           & 0.1004           & 0.4588          & 0.5823            & 0.9545               & 0.0293           & 0.9281            & 0.7301            & 0.7431            & 0.5985          \\
cryptography                & 0.9143           & 0.6395           & 0.4864          & 0.6911            & 0.7596               & 0.6896           & 0.0717            & 0.6523            & 0.0297            & 0.1090          \\
dogecoin                    & 0.4168           & 0.0745           & 0.8324          & \textbf{0.0033}   & 0.5577               & 0.7827           & 0.3509            & 0.1329            & 0.5748            & 0.0572          \\
digital gold                & 0.1681           & 0.6520           & \textbf{0.0052} & 0.6572            & 0.2344               & 0.1247           & 0.3808            & 0.7552            & 0.1092            & 0.0133          \\
Bitcoin                     & 0.8074           & \textbf{0.0000}  & 0.1341          & 0.0637            & 0.6245               & 0.9207           & 0.0225            & 0.3835            & 0.5476            & 0.1504          \\
NFT                         & 0.8955           & 0.4895           & 0.9963          & 0.3886            & 0.6583               & 0.9873           & 0.3122            & 0.5264            & 0.6557            & 0.7778          \\
ripple                      & 0.8699           & 0.6457           & 0.2431          & 0.6111            & 0.3368               & 0.6968           & 0.6324            & 0.5787            & 0.6279            & 0.4275          \\
private key                 & 0.9722           & 0.3132           & 0.2969          & 0.4290            & 0.6356               & 0.5147           & 0.1474            & 0.0519            & 0.6744            & 0.8022          \\
peercoin                    & 0.1326           & 0.0460           & 0.1669          & 0.1729            & 0.5522               & 0.5884           & 0.5705            & 0.4234            & 0.7349            & 0.7582          \\
namecoin                    & 0.3438           & 0.6583           & 0.6245          & 0.1543            & 0.9187               & 0.4374           & 0.0651            & 0.4058            & 0.7169            & 0.1354          \\
satoshi                     & 0.4650           & 0.0235           & 0.5989          & 0.8808            & 0.3549               & 0.2758           & 0.3560            & 0.4398            & 0.5796            & 0.7183          \\
hot wallet                  & 0.6852           & 0.7534           & 0.0276          & 0.0691            & 0.8563               & 0.5838           & 0.6089            & 0.8984            & 0.9160            & 0.4712          \\
blockchain                  & 0.2278           & 0.9625           & 0.6401          & 0.2790            & 0.0117               & 0.6034           & 0.3028            & 0.8676            & 0.1049            & 0.2246          \\
ethereum                    & 0.6211           & 0.0292           & 0.6188          & 0.0367            & 0.8557               & 0.6875           & 0.0681            & 0.4486            & 0.2699            & 0.0974          \\
zen coin                    & 0.9192           & 0.8343           & 0.9160          & 0.9038            & 0.7183               & 0.1420           & 0.9143            & 0.7355            & 0.7321            & 0.6497          \\
coinbase                    & 0.8630           & 0.0119           & 0.5855          & \textbf{0.0071}   & 0.5251               & 0.9299           & 0.1896            & 0.1725            & 0.1492            & 0.0325          \\
BTC                         & 0.8833           & \textbf{0.0000}  & 0.4303          & 0.7704            & 0.3939               & 0.8184           & 0.0366            & 0.2505            & 0.2538            & 0.7446          \\
hard fork                   & 0.1036           & 0.7880           & 0.8580          & \textbf{0.0009}   & 0.6449               & 0.6969           & 0.7318            & 0.6394            & 0.7263            & 0.1168          \\
stablecoin                  & 0.2091           & 0.8364           & 0.4060          & 0.8699            & 0.3242               & 0.4720           & 0.0347            & 0.0490            & 0.4366            & 0.5818          \\
cold storage                & 0.9267           & 0.6904           & 0.8710          & 0.6980            & 0.7578               & 0.3718           & 0.9898            & 0.4618            & 0.5048            & 0.4033          \\
coin                        & 0.4056           & 0.1404           & 0.6030          & 0.9604            & 0.1383               & 0.3260           & 0.2258            & 0.8750            & 0.2673            & 0.2704          \\
hash                        & 0.7988           & 0.0937           & 0.2399          & 0.9853            & 0.6009               & 0.2439           & 0.3649            & 0.8882            & 0.6758            & 0.6788          \\
tether                      & 0.9119           & 0.4673           & 0.1943          & 0.5010            & 0.4542               & 0.8451           & 0.3022            & 0.6103            & 0.3950            & 0.1726          \\
litecoin                    & 0.9384           & 0.0692           & 0.3162          & 0.1120            & 0.4842               & 0.2935           & 0.5608            & 0.9612            & 0.4839            & 0.2853          \\
enjin coin                  & 0.8819           & 0.5395           & 0.1981          & 0.7181            & 0.8608               & 0.3886           & 0.0201            & 0.3699            & 0.7663            & 0.4399          \\
hashing                     & 0.7452           & 0.0403           & 0.9332          & 0.5851            & 0.9850               & 0.1087           & 0.4256            & 0.1738            & 0.5225            & 0.3914          \\
token                       & 0.0315           & 0.1228           & 0.0745          & 0.9948            & 0.1451               & 0.8357           & 0.7580            & 0.4954            & 0.5339            & 0.0574          \\
crypto                      & 0.2607           & \textbf{0.0005}  & 0.0239          & 0.0101            & 0.5988               & 0.9822           & 0.0867            & 0.4062            & 0.0304            & 0.3774          \\
ICO                         & 0.9978           & 0.7753           & 0.5554          & 0.3998            & 0.1844               & 0.2904           & 0.7830            & 0.8071            & 0.2732            & 0.2678          \\
coin market                 & 0.5826           & \textbf{0.0077}  & 0.5031          & 0.9928            & 0.8572               & 0.3036           & 0.4974            & 0.8567            & 0.4990            & 0.4338          \\
soft fork                   & 0.2140           & 0.9266           & 0.6588          & 0.7629            & 0.9294               & 0.1710           & 0.4798            & 0.3797            & 0.2247            & 0.2018          \\
block producer              & 0.8742           & 0.0726           & 0.5215          & 0.2799            & 0.4695               & 0.1274           & 0.4518            & 0.1755            & 0.5614            & 0.1294          \\
distributed ledger          & 0.1064           & 0.6772           & 0.4766          & 0.4593            & 0.1056               & 0.2178           & 0.7169            & 0.4729            & 0.8338            & 0.3056          \\
digital fiat                & 0.6758           & 0.4717           & 0.8820          & 0.9073            & 0.8926               & 0.3863           & 0.3462            & 0.3646            & 0.2683            & 0.2560          \\
consensus                   & 0.5748           & 0.8610           & 0.0248          & 0.4499            & 0.1213               & 0.6976           & \textbf{0.0001}   & 0.1220            & 0.0772            & 0.9201          \\
BTC-USD\_vol                & 0.6931           & 0.1518           & 0.1349          & 0.9072            & 0.7198               & 0.9983           & \textbf{0.0068}   & 0.9658            & 0.4401            & 0.3173          \\
ETH-USD\_vol                & 0.8302           & 0.0182           & 0.2177          & 0.5473            & 0.5977               & 0.9582           & 0.0103            & 0.7157            & 0.7731            & 0.1990          \\
USDT-USD\_vol               & 0.9718           & 0.2938           & 0.0612          & 0.5901            & 0.3968               & 0.9538           & 0.0439            & 0.8909            & 0.5632            & 0.4050          \\
BNB-USD\_vol                & 0.8680           & 0.1704           & 0.2059          & 0.7025            & 0.4337               & \textbf{0.0041}  & 0.6148            & 0.7409            & 0.6974            & 0.9243          \\
LTC-USD\_vol                & 0.9636           & 0.1657           & 0.6467          & 0.0371            & 0.2931               & 0.7253           & 0.1516            & 0.5530            & 0.1434            & 0.5489          \\
ENJ-USD\_vol                & 0.3668           & 0.4262           & 0.5411          & 0.7350            & 0.1881               & 0.6001           & 0.9762            & 0.2902            & 0.3975            & 0.4950          \\
ZEN-USD\_vol                & 0.9999           & 0.0847           & 0.0284          & 0.1301            & 0.7877               & 0.3356           & 0.5069            & 0.8507            & 0.9785            & 0.9577          \\
NMC-USD\_vol                & 0.7818           & 0.4246           & 0.4463          & 0.8682            & 0.1124               & 0.6994           & 0.2935            & 0.6654            & 0.3983            & 0.6516          \\
PPC-USD\_vol                & 0.6717           & 0.0425           & 0.3916          & 0.0999            & 0.7151               & 0.4440           & 0.3600            & 0.7065            & 0.2625            & 0.8024          \\
FTC-USD\_vol                & 0.8425           & 0.6522           & 0.1023          & 0.0854            & 0.6400               & 0.4782           & 0.0640            & 0.1272            & 0.7489            & 0.6695          \\ \hline
\end{tabular}
\end{adjustbox}
\end{table}
\begin{table}[]
\caption{$\textit{p-values}$ of Post-double-selection Granger causality LM test from Reddit sentiment to other variables }
\label{GCtable}
\begin{adjustbox}{ max width=1\textwidth, center}
\begin{tabular}{lllllllllll}
\hline
Gcto                 GCfrom & binance\_reddit & bitcoin\_reddit & enjin\_reddit & ethereum\_reddit & feathercoin\_reddit & horizen\_reddit & litecoin\_reddit & namecoin\_reddit & peercoin\_reddit & tether\_reddit  \\ \hline
BTC-USD                     & 0.6490          & 0.5749          & 0.7952        & 0.7392           & 0.0162              & 0.9303          & 0.9640           & 0.3521           & 0.4535           & 0.6762          \\
ETH-USD                     & 0.6510          & 0.4152          & 0.2781        & 0.2011           & 0.0758              & 0.9016          & 0.4627           & 0.5559           & 0.8040           & 0.5978          \\
USDT-USD                    & 0.0434          & 0.4024          & 0.5510        & 0.2960           & 0.9342              & 0.7445          & 0.9259           & 0.4391           & 0.7519           & 0.8106          \\
BNB-USD                     & 0.5846          & 0.1236          & 0.8067        & 0.5063           & \textbf{0.0003}     & 0.3475          & 0.9164           & 0.7684           & 0.0287           & 0.4639          \\
LTC-USD                     & 0.5551          & 0.8095          & 0.5847        & 0.6245           & \textbf{0.0009}     & 0.5876          & 0.6108           & 0.7981           & 0.2549           & 0.8514          \\
ENJ-USD                     & 0.1720          & 0.5678          & 0.5899        & 0.3021           & 0.0248              & 0.8287          & 0.8361           & 0.2641           & 0.3282           & 0.0459          \\
ZEN-USD                     & 0.7415          & 0.0485          & 0.3935        & 0.4864           & 0.0475              & 0.1194          & 0.2606           & 0.2864           & 0.5880           & 0.1124          \\
NMC-USD                     & 0.7376          & 0.2239          & 0.0255        & 0.7927           & 0.6858              & 0.2549          & 0.0966           & 0.2419           & 0.3208           & 0.1889          \\
PPC-USD                     & 0.9175          & 0.8382          & 0.0492        & 0.3691           & \textbf{0.0011}     & 0.7565          & 0.9215           & 0.5669           & 0.1147           & 0.6673          \\
FTC-USD                     & 0.4611          & 0.0510          & 0.2478        & 0.5661           & 0.0722              & 0.8019          & 0.5361           & 0.4193           & 0.7658           & 0.1904          \\
binance\_twitter            & 0.3016          & 0.5025          & 0.0191        & 0.1006           & 0.6707              & 0.2966          & 0.6769           & 0.6569           & 0.0541           & 0.3627          \\
bitcoin\_twitter            & 0.6883          & 0.2564          & 0.2780        & 0.3883           & 0.2888              & 0.8810          & 0.6937           & 0.3852           & 0.4587           & 0.0527          \\
enjin\_twitter              & 0.0217          & 0.3034          & 0.2111        & 0.5200           & 0.9675              & 0.5887          & 0.4704           & 0.0240           & 0.0668           & 0.9908          \\
ethereum\_twitter           & 0.5575          & 0.4115          & 0.8155        & 0.5059           & 0.5993              & 0.3075          & 0.6996           & 0.8148           & 0.5905           & 0.4012          \\
feathercoin\_twitter        & 0.5076          & 0.4717          & 0.9325        & 0.1416           & 0.3479              & 0.2080          & \textbf{0.0071}  & 0.4969           & 0.1132           & 0.4819          \\
horizen\_twitter            & 0.7657          & 0.9124          & 0.2826        & 0.7095           & 0.0872              & 0.8510          & 0.9174           & 0.6230           & 0.5977           & 0.9491          \\
litecoin\_twitter           & 0.0804          & 0.3251          & 0.7399        & 0.9148           & 0.1134              & 0.8502          & 0.1364           & 0.1193           & 0.7239           & 0.5644          \\
namecoin\_twitter           & 0.0762          & 0.2868          & 0.5301        & 0.8203           & 0.8946              & 0.4796          & 0.1818           & 0.7366           & 0.0278           & 0.4874          \\
peercoin\_twitter           & 0.9871          & 0.3294          & 0.3634        & 0.9812           & 0.8985              & 0.2443          & 0.4211           & 0.8195           & 0.7866           & 0.5209          \\
tether\_twitter             & 0.3473          & 0.5107          & 0.3276        & 0.5045           & 0.4230              & 0.6701          & 0.3549           & 0.5638           & 0.7395           & 0.4016          \\
binance\_reddit             & NA              & \textbf{0.0074} & 0.5982        & 0.0940           & 0.4055              & 0.1659          & 0.0168           & 0.9259           & 0.2877           & 0.9722          \\
bitcoin\_reddit             & 0.8617          & NA              & 0.0396        & \textbf{0.0035}  & 0.0170              & 0.2755          & 0.0915           & 0.3345           & 0.7027           & 0.8080          \\
enjin\_reddit               & 0.6899          & 0.0124          & NA            & 0.8998           & 0.1595              & 0.4637          & 0.5484           & 0.0188           & 0.1576           & 0.9336          \\
ethereum\_reddit            & \textbf{0.0034} & \textbf{0.0001} & 0.1593        & NA               & 0.9247              & 0.1884          & 0.0469           & 0.0136           & 0.5647           & 0.5692          \\
feathercoin\_reddit         & 0.6173          & 0.9384          & 0.0695        & 0.0457           & NA                  & 0.2614          & 0.4184           & 0.3888           & 0.7616           & 0.2512          \\
horizen\_reddit             & 0.8246          & 0.1495          & 0.9839        & \textbf{0.0070}  & 0.9802              & NA              & 0.6683           & 0.4089           & 0.4249           & 0.4498          \\
litecoin\_reddit            & 0.0658          & \textbf{0.0002} & 0.5461        & 0.4101           & 0.9522              & 0.2247          & NA               & 0.6164           & 0.1079           & 0.0194          \\
namecoin\_reddit            & 0.0102          & 0.6317          & 0.1763        & 0.3161           & 0.6554              & 0.1168          & 0.9148           & NA               & 0.0164           & 0.1238          \\
peercoin\_reddit            & 0.6017          & 0.8561          & 0.5785        & 0.4710           & 0.3278              & 0.9843          & 0.6041           & 0.1234           & NA               & 0.1801          \\
tether\_reddit              & 0.6047          & 0.9786          & 0.5541        & 0.7625           & 0.3930              & 0.7942          & 0.3073           & \textbf{0.0042}  & 0.3508           & NA              \\
binance coin                & 0.2994          & 0.3339          & 0.6540        & 0.2915           & 0.2537              & 0.4130          & \textbf{0.0024}  & 0.5475           & 0.0346           & 0.6398          \\
crypto crash                & 0.9829          & 0.7047          & 0.7110        & 0.8784           & \textbf{0.0001}     & 0.2687          & 0.9814           & 0.8713           & 0.4930           & 0.2463          \\
miner                       & 0.2503          & 0.8835          & 0.8058        & 0.8159           & 0.4789              & 0.3474          & 0.4953           & 0.1988           & 0.4143           & 0.0389          \\
USD coin                    & 0.7764          & 0.0889          & 0.0187        & 0.1321           & 0.8106              & 0.4035          & 0.8317           & 0.2448           & 0.0216           & 0.5041          \\
cardano                     & 0.7831          & 0.9302          & 0.3779        & 0.3630           & \textbf{0.0034}     & 0.2814          & 0.9166           & 0.1577           & 0.8480           & 0.6328          \\
cryptocurrency              & 0.9764          & 0.1748          & 0.2151        & 0.0113           & 0.0215              & 0.4712          & 0.4463           & 0.4955           & 0.9182           & 0.7650          \\
minted                      & 0.9614          & 0.0781          & 0.4962        & 0.2967           & 0.3863              & 0.1412          & 0.6220           & 0.5780           & 0.6831           & 0.6189          \\
solana                      & 0.6419          & 0.6476          & 0.5900        & 0.6248           & 0.8410              & 0.4601          & 0.6212           & 0.7741           & 0.2162           & 0.6842          \\
feathercoin                 & 0.7856          & \textbf{0.0080} & 0.7514        & 0.9825           & 0.9420              & 0.9964          & 0.7204           & 0.1433           & 0.3573           & 0.0417          \\
cryptography                & 0.2454          & 0.9760          & 0.5892        & 0.5569           & 0.4638              & 0.8787          & 0.1167           & 0.3347           & 0.9007           & 0.4522          \\
dogecoin                    & 0.2900          & \textbf{0.0004} & 0.2622        & \textbf{0.0064}  & 0.0280              & 0.8273          & 0.0441           & 0.9780           & 0.7086           & 0.9650          \\
digital gold                & 0.7365          & 0.1447          & 0.8024        & 0.8288           & 0.7157              & 0.7786          & 0.8151           & 0.9571           & 0.7435           & 0.1264          \\
Bitcoin                     & 0.5919          & 0.5289          & 0.0401        & 0.0181           & 0.2721              & 0.0699          & 0.7863           & 0.7879           & 0.7983           & 0.8720          \\
NFT                         & 0.9891          & 0.0447          & 0.9048        & 0.9869           & \textbf{0.0040}     & 0.2173          & 0.9988           & 0.7431           & 0.5836           & 0.3398          \\
ripple                      & 0.6763          & 0.7278          & 0.5541        & 0.8294           & 0.5933              & 0.2928          & 0.6975           & 0.7715           & 0.2012           & 0.5337          \\
private key                 & 0.4332          & 0.9124          & 0.4249        & 0.6011           & 0.5305              & 0.3144          & 0.6888           & \textbf{0.0000}  & 0.8677           & 0.2813          \\
peercoin                    & 0.5794          & 0.7239          & 0.1405        & 0.7151           & 0.4742              & 0.2392          & 0.9847           & 0.8850           & 0.0499           & 0.4968          \\
namecoin                    & 0.1184          & 0.0184          & 0.6779        & 0.0982           & 0.1126              & 0.1430          & 0.9001           & 0.4028           & 0.1248           & 0.2541          \\
satoshi                     & 0.7215          & 0.3168          & 0.3788        & 0.7878           & 0.1415              & 0.6646          & 0.7774           & 0.0922           & \textbf{0.0053}  & 0.0580          \\
hot wallet                  & 0.0707          & 0.7975          & 0.3375        & 0.7932           & 0.9116              & 0.3362          & 0.8039           & 0.2399           & 0.7290           & 0.7211          \\
blockchain                  & 0.3047          & 0.0112          & 0.3233        & 0.1376           & 0.0666              & 0.5441          & 0.3690           & 0.7781           & 0.2715           & 0.8976          \\
ethereum                    & 0.8876          & 0.1368          & 0.0966        & 0.3215           & \textbf{0.0000}     & 0.0833          & 0.6417           & 0.9604           & 0.4382           & 0.9158          \\
zen coin                    & 0.5206          & 0.0881          & 0.1203        & 0.0757           & 0.9548              & 0.8401          & 0.0672           & 0.8691           & 0.9197           & 0.0894          \\
coinbase                    & 0.4432          & 0.0207          & 0.0861        & \textbf{0.0038}  & 0.1051              & 0.4156          & 0.3609           & 0.4592           & 0.8001           & 0.1457          \\
BTC                         & 0.8161          & 0.9790          & 0.4817        & 0.8776           & 0.5372              & \textbf{0.0069} & 0.6368           & 0.7545           & 0.7117           & 0.8104          \\
hard fork                   & 0.4211          & 0.9022          & 0.4761        & 0.2127           & 0.8809              & 0.9924          & 0.7434           & 0.8987           & 0.7459           & 0.5645          \\
stablecoin                  & 0.8152          & 0.0392          & 0.0988        & 0.6329           & 0.6643              & 0.1781          & 0.4136           & 0.8372           & 0.6247           & 0.5707          \\
cold storage                & 0.4598          & 0.7226          & 0.3689        & 0.5670           & 0.4642              & 0.2388          & 0.2349           & 0.5179           & 0.9288           & 0.8961          \\
coin                        & 0.5338          & 0.4550          & 0.7080        & 0.2174           & 0.2796              & 0.4005          & 0.5509           & 0.2503           & 0.6287           & 0.3292          \\
hash                        & 0.4706          & 0.1316          & 0.9177        & 0.2431           & 0.8156              & 0.8129          & 0.4942           & 0.4698           & 0.9315           & 0.1682          \\
tether                      & 0.7797          & 0.9060          & 0.2313        & 0.3477           & 0.1724              & 0.4384          & 0.6333           & 0.3970           & 0.5414           & 0.6777          \\
litecoin                    & 0.2674          & 0.6376          & 0.7509        & 0.8048           & 0.1171              & 0.7526          & 0.2450           & 0.3120           & 0.9666           & 0.6412          \\
enjin coin                  & 0.8524          & 0.2545          & 0.8180        & 0.0535           & 0.0946              & 0.8839          & 0.1952           & 0.1628           & 0.9111           & 0.2115          \\
hashing                     & 0.9460          & 0.2187          & 0.0746        & 0.8880           & 0.3560              & 0.6763          & 0.7199           & 0.3911           & 0.8885           & 0.6406          \\
token                       & 0.5129          & 0.9908          & 0.8293        & 0.6380           & 0.4843              & 0.3687          & 0.2607           & 0.3970           & 0.7820           & 0.5039          \\
crypto                      & 0.6042          & 0.0972          & 0.1664        & 0.0671           & 0.0373              & 0.5587          & 0.3189           & 0.5801           & 0.5257           & 0.9642          \\
ICO                         & 0.1847          & 0.8678          & 0.1832        & 0.3347           & 0.5071              & 0.6415          & 0.8833           & 0.0879           & 0.8129           & 0.8119          \\
coin market                 & 0.5790          & 0.8545          & 0.3920        & 0.8424           & 0.5729              & 0.6280          & 0.6597           & 0.1108           & 0.9075           & 0.6824          \\
soft fork                   & 0.4795          & 0.5505          & 0.5478        & 0.6459           & 0.0677              & 0.1177          & 0.4669           & 0.6815           & 0.7157           & \textbf{0.0018} \\
block producer              & 0.3202          & 0.3623          & 0.7920        & 0.0372           & 0.6525              & 0.7791          & 0.9513           & 0.3789           & 0.0863           & 0.4046          \\
distributed ledger          & 0.2198          & 0.1659          & 0.9087        & 0.3150           & 0.5804              & 0.9496          & 0.5736           & 0.1056           & 0.1256           & 0.1218          \\
digital fiat                & 0.3504          & 0.8585          & 0.1294        & 0.4523           & 0.7172              & 0.8056          & 0.1239           & 0.6280           & 0.3468           & 0.1420          \\
consensus                   & 0.9329          & 0.5980          & 0.3054        & 0.6781           & 0.4437              & 0.3079          & 0.6963           & 0.2267           & 0.0539           & 0.0531          \\
BTC-USD\_vol                & 0.9794          & 0.1726          & 0.0586        & 0.0796           & 0.3739              & 0.2771          & 0.7123           & 0.7395           & 0.8830           & 0.0815          \\
ETH-USD\_vol                & 0.8812          & 0.4840          & 0.0841        & 0.0494           & 0.3595              & 0.1572          & 0.9486           & 0.9443           & 0.5026           & 0.1569          \\
USDT-USD\_vol               & 0.8782          & 0.0256          & 0.0296        & 0.0637           & 0.2016              & 0.5154          & 0.9347           & 0.5083           & 0.8106           & 0.0415          \\
BNB-USD\_vol                & 0.1647          & 0.3737          & 0.1669        & 0.4455           & 0.9611              & 0.9427          & 0.7012           & 0.9340           & 0.8812           & 0.9726          \\
LTC-USD\_vol                & 0.6387          & 0.0244          & 0.6301        & 0.1707           & 0.0822              & 0.4796          & 0.5236           & 0.8860           & 0.4430           & 0.2986          \\
ENJ-USD\_vol                & 0.6992          & 0.6197          & 0.6292        & 0.5137           & 0.8576              & 0.7347          & 0.9672           & 0.0995           & 0.1144           & 0.2575          \\
ZEN-USD\_vol                & 0.4637          & 0.0505          & 0.3007        & 0.3367           & 0.6024              & 0.2327          & 0.1100           & 0.3214           & 0.3580           & 0.9632          \\
NMC-USD\_vol                & \textbf{0.0035} & 0.0837          & 0.4043        & 0.0807           & 0.7069              & 0.2048          & 0.4870           & 0.5559           & 0.9977           & 0.2749          \\
PPC-USD\_vol                & 0.9681          & 0.2028          & 0.7239        & 0.5624           & 0.0594              & 0.7154          & 0.9853           & 0.0296           & 0.9202           & 0.4822          \\
FTC-USD\_vol                & 0.9126          & 0.7379          & 0.8872        & 0.9207           & 0.9066              & 0.5657          & 0.8162           & 0.8388           & 0.6192           & 0.7975          \\ \hline
\end{tabular}
\end{adjustbox}
\end{table}
\begin{table}[]
\caption{$\textit{p-values}$ of Post-double-selection Granger causality LM test from Volume to other variables }
\label{tab:my-table}
\begin{adjustbox}{ max width=1\textwidth, center}
\begin{tabular}{lllllllllll}
\hline
Gcto                 GCfrom & BTC-USD\_vol    & ETH-USD\_vol    & USDT-USD\_vol   & BNB-USD\_vol    & LTC-USD\_vol    & ENJ-USD\_vol    & ZEN-USD\_vol    & NMC-USD\_vol    & PPC-USD\_vol    & FTC-USD\_vol    \\ \hline
BTC-USD                     & 0.4285          & 0.2040          & 0.1035          & \textbf{0.0076} & 0.1270          & 0.9440          & 0.9111          & 0.5741          & 0.3852          & 0.7540          \\
ETH-USD                     & 0.3751          & 0.3762          & 0.1812          & 0.6225          & 0.4007          & 0.8975          & 0.6259          & 0.7745          & 0.2047          & 0.9593          \\
USDT-USD                    & 0.0394          & 0.3672          & 0.0245          & 0.4170          & 0.1595          & 0.5817          & 0.7504          & 0.0224          & 0.4298          & 0.4827          \\
BNB-USD                     & 0.1276          & 0.0598          & 0.0361          & 0.3857          & 0.0943          & 0.4642          & 0.9655          & \textbf{0.0021} & 0.1347          & 0.8454          \\
LTC-USD                     & 0.0460          & 0.3050          & 0.0355          & 0.0965          & 0.1551          & 0.9016          & 0.6767          & 0.1582          & 0.1387          & 0.6775          \\
ENJ-USD                     & 0.0879          & 0.0469          & \textbf{0.0070} & 0.0177          & 0.0309          & 0.6038          & 0.4753          & 0.7882          & 0.0925          & 0.5563          \\
ZEN-USD                     & 0.0526          & 0.0218          & 0.0214          & 0.2155          & 0.0925          & 0.6199          & 0.0741          & 0.3548          & 0.4560          & 0.2485          \\
NMC-USD                     & 0.3271          & 0.5458          & 0.5675          & 0.1612          & 0.6613          & 0.3852          & 0.9786          & 0.0133          & 0.8069          & 0.4959          \\
PPC-USD                     & 0.4082          & 0.1950          & 0.4555          & 0.2910          & 0.7178          & 0.6623          & 0.2226          & 0.6966          & 0.1463          & 0.7791          \\
FTC-USD                     & 0.0188          & \textbf{0.0012} & 0.0126          & 0.0823          & 0.0826          & 0.7961          & 0.1825          & 0.8718          & 0.8897          & 0.4579          \\
binance\_twitter            & 0.6496          & 0.6694          & 0.5650          & 0.8522          & 0.7673          & 0.9552          & 0.7964          & 0.5226          & 0.3761          & 0.1815          \\
bitcoin\_twitter            & 0.5083          & 0.8032          & 0.2585          & 0.7420          & 0.3441          & 0.1034          & 0.4094          & 0.3326          & 0.0885          & 0.6875          \\
enjin\_twitter              & 0.8471          & 0.6550          & 0.6885          & 0.8965          & 0.7260          & 0.1598          & 0.2833          & 0.8763          & 0.8123          & 0.3686          \\
ethereum\_twitter           & 0.9305          & 0.2036          & 0.9729          & 0.6032          & 0.6708          & 0.2566          & 0.8449          & 0.1768          & 0.0139          & 0.7689          \\
feathercoin\_twitter        & 0.2271          & 0.3822          & 0.2634          & 0.4842          & 0.8401          & 0.7611          & 0.0718          & 0.3284          & 0.9371          & 0.1901          \\
horizen\_twitter            & 0.0342          & 0.0624          & \textbf{0.0049} & \textbf{0.0011} & 0.0943          & 0.2569          & \textbf{0.0022} & 0.3027          & 0.7935          & 0.7720          \\
litecoin\_twitter           & 0.0732          & 0.4514          & 0.5763          & 0.7475          & 0.8273          & 0.4747          & 0.9866          & 0.6513          & 0.4298          & 0.9762          \\
namecoin\_twitter           & 0.5226          & 0.6241          & 0.7155          & 0.8615          & 0.8546          & 0.1329          & 0.8688          & 0.5213          & 0.6945          & 0.7958          \\
peercoin\_twitter           & 0.1182          & 0.1154          & 0.2793          & 0.2588          & 0.2345          & 0.2260          & 0.9807          & 0.2259          & 0.3138          & 0.7869          \\
tether\_twitter             & 0.5719          & 0.7046          & 0.8761          & 0.5646          & 0.9422          & 0.4827          & 0.8046          & 0.2573          & 0.2637          & 0.9184          \\
binance\_reddit             & 0.8505          & 0.6867          & 0.6462          & 0.6421          & 0.9327          & 0.3752          & 0.8229          & 0.5030          & 0.8442          & \textbf{0.0008} \\
bitcoin\_reddit             & 0.0523          & 0.7124          & 0.7319          & 0.8014          & 0.6673          & 0.0788          & 0.9135          & 0.7375          & 0.0729          & 0.4891          \\
enjin\_reddit               & 0.2345          & 0.7225          & 0.3328          & 0.6972          & 0.7382          & 0.9602          & 0.2970          & 0.1516          & 0.6985          & 0.9444          \\
ethereum\_reddit            & 0.0142          & 0.4270          & 0.5509          & 0.9805          & 0.4062          & 0.6279          & 0.6412          & 0.1812          & 0.6234          & 0.3045          \\
feathercoin\_reddit         & 0.0478          & 0.0696          & 0.2066          & 0.9613          & 0.4141          & 0.9436          & 0.0151          & 0.8839          & 0.3658          & 0.1721          \\
horizen\_reddit             & 0.1343          & 0.2394          & 0.2839          & 0.9423          & 0.0954          & 0.3787          & 0.2212          & 0.6524          & 0.7528          & 0.7723          \\
litecoin\_reddit            & 0.5617          & 0.7762          & 0.8381          & 0.9978          & 0.8776          & 0.0338          & 0.6787          & 0.1220          & 0.7442          & 0.3543          \\
namecoin\_reddit            & 0.3859          & 0.6345          & 0.1390          & 0.6970          & 0.3665          & 0.3975          & 0.4890          & 0.2765          & 0.7093          & 0.8338          \\
peercoin\_reddit            & 0.3535          & 0.2973          & 0.1909          & 0.9201          & 0.7866          & 0.0812          & 0.1712          & 0.0791          & 0.1725          & 0.2772          \\
tether\_reddit              & 0.7572          & 0.4570          & 0.8763          & 0.9524          & 0.5835          & 0.2375          & 0.6782          & 0.5091          & 0.1027          & 0.8731          \\
binance coin                & 0.4657          & 0.3887          & 0.2116          & 0.5002          & 0.0242          & 0.9713          & 0.7909          & 0.0872          & 0.5812          & 0.8036          \\
crypto crash                & 0.3252          & \textbf{0.0075} & 0.1114          & 0.7650          & 0.0127          & 0.5786          & 0.6972          & 0.4862          & \textbf{0.0002} & 0.7896          \\
miner                       & 0.3695          & 0.0756          & 0.6606          & 0.9977          & 0.2836          & 0.9908          & 0.0749          & 0.9332          & 0.5673          & 0.2072          \\
USD coin                    & \textbf{0.0040} & 0.0261          & \textbf{0.0020} & 0.2067          & \textbf{0.0039} & 0.1276          & 0.8474          & 0.1049          & 0.3549          & 0.0304          \\
cardano                     & 0.0612          & 0.1818          & 0.0433          & 0.7445          & 0.0654          & 0.6280          & 0.0623          & 0.8432          & 0.7648          & 0.6225          \\
cryptocurrency              & 0.0443          & \textbf{0.0006} & \textbf{0.0023} & 0.4431          & 0.0310          & 0.7356          & 0.3900          & 0.4141          & 0.4391          & 0.8441          \\
minted                      & 0.1533          & 0.1233          & 0.1424          & 0.1922          & 0.4464          & 0.4659          & 0.1948          & 0.3541          & 0.9067          & 0.8170          \\
solana                      & 0.4350          & 0.3274          & 0.6674          & 0.6826          & 0.6486          & 0.4298          & 0.9597          & 0.8354          & 0.8515          & 0.1101          \\
feathercoin                 & 0.9341          & 0.7550          & 0.8798          & \textbf{0.0061} & 0.3923          & 0.8571          & 0.3049          & 0.1966          & 0.3088          & 0.6064          \\
cryptography                & \textbf{0.0001} & \textbf{0.0049} & \textbf{0.0000} & 0.3210          & 0.0774          & 0.8986          & 0.0526          & 0.7511          & 0.3581          & 0.2711          \\
dogecoin                    & 0.0152          & \textbf{0.0048} & 0.0183          & 0.4301          & 0.0181          & 0.3140          & 0.2675          & 0.4381          & 0.4478          & 0.8318          \\
digital gold                & 0.0106          & 0.0612          & 0.0130          & 0.0520          & \textbf{0.0028} & 0.8655          & 0.8759          & 0.6721          & 0.3912          & 0.3434          \\
Bitcoin                     & \textbf{0.0008} & \textbf{0.0002} & \textbf{0.0003} & 0.9775          & 0.0874          & 0.5390          & 0.6651          & 0.9751          & 0.6542          & 0.7775          \\
NFT                         & 0.0168          & 0.0242          & 0.0154          & 0.7478          & 0.0179          & 0.7931          & 0.5051          & 0.9631          & 0.1214          & 0.2062          \\
ripple                      & 0.1916          & 0.2840          & 0.4704          & 0.1208          & 0.4165          & 0.6319          & 0.1985          & 0.5323          & 0.1049          & 0.4354          \\
private key                 & 0.0172          & 0.0119          & \textbf{0.0029} & 0.3754          & 0.1041          & 0.1045          & 0.6450          & 0.8844          & 0.1006          & 0.0710          \\
peercoin                    & 0.6156          & 0.4333          & 0.6224          & 0.7550          & 0.0404          & 0.0951          & 0.6624          & 0.8133          & \textbf{0.0000} & 0.8569          \\
namecoin                    & 0.2319          & 0.0130          & 0.3100          & 0.6094          & 0.2251          & 0.0816          & 0.1816          & 0.6161          & 0.6859          & 0.4534          \\
satoshi                     & 0.0145          & 0.0904          & \textbf{0.0040} & 0.1417          & \textbf{0.0019} & 0.5589          & 0.1936          & 0.2889          & 0.4196          & 0.0616          \\
hot wallet                  & 0.6243          & 0.3386          & 0.6166          & 0.0184          & 0.5466          & 0.9157          & 0.4923          & 0.9377          & 0.1872          & 0.5798          \\
blockchain                  & 0.3693          & 0.3014          & 0.2509          & 0.8575          & 0.2406          & 0.5028          & 0.0812          & 0.6919          & 0.2843          & 0.7424          \\
ethereum                    & \textbf{0.0023} & \textbf{0.0001} & \textbf{0.0000} & 0.8153          & \textbf{0.0028} & 0.7266          & \textbf{0.0098} & 0.3730          & 0.5688          & 0.5747          \\
zen coin                    & 0.0860          & 0.4967          & 0.1826          & 0.9928          & 0.1747          & 0.5511          & 0.7401          & 0.8868          & 0.9089          & 0.5085          \\
coinbase                    & \textbf{0.0002} & \textbf{0.0000} & \textbf{0.0001} & 0.7880          & 0.1245          & 0.4326          & \textbf{0.0052} & 0.5600          & 0.1244          & 0.3437          \\
BTC                         & \textbf{0.0000} & \textbf{0.0000} & \textbf{0.0000} & 0.7923          & \textbf{0.0000} & 0.5018          & 0.1289          & 0.5378          & 0.5574          & 0.8746          \\
hard fork                   & 0.7006          & 0.1778          & 0.3790          & 0.4508          & 0.1438          & 0.3219          & 0.4376          & 0.5364          & 0.1437          & 0.8365          \\
stablecoin                  & 0.0649          & 0.1285          & 0.0331          & 0.1504          & 0.1008          & 0.6498          & 0.1073          & 0.1693          & 0.4644          & 0.5789          \\
cold storage                & \textbf{0.0000} & \textbf{0.0000} & \textbf{0.0000} & 0.5024          & \textbf{0.0013} & 0.5624          & 0.0174          & 0.6038          & 0.6207          & 0.8655          \\
coin                        & \textbf{0.0000} & \textbf{0.0000} & \textbf{0.0000} & 0.0474          & \textbf{0.0011} & 0.0240          & \textbf{0.0082} & 0.3114          & 0.7081          & 0.3721          \\
hash                        & \textbf{0.0007} & 0.0157          & 0.0210          & 0.4702          & 0.2266          & 0.5799          & 0.9762          & 0.0574          & 0.2057          & 0.7810          \\
tether                      & 0.1546          & 0.0189          & 0.0421          & 0.9706          & 0.0746          & 0.3558          & 0.5623          & 0.9917          & 0.3505          & 0.5671          \\
litecoin                    & 0.0626          & \textbf{0.0011} & \textbf{0.0006} & \textbf{0.0000} & \textbf{0.0055} & 0.0207          & 0.1481          & 0.4020          & 0.9960          & 0.9071          \\
enjin coin                  & 0.0596          & 0.3111          & 0.3243          & 0.1006          & 0.1710          & \textbf{0.0000} & 0.6518          & 0.0559          & 0.9434          & 0.2782          \\
hashing                     & \textbf{0.0000} & \textbf{0.0000} & \textbf{0.0000} & 0.4573          & \textbf{0.0046} & 0.3344          & 0.4514          & 0.1547          & 0.3340          & 0.5922          \\
token                       & \textbf{0.0002} & \textbf{0.0058} & \textbf{0.0010} & 0.8487          & 0.1697          & 0.0991          & 0.2591          & 0.3528          & 0.8180          & 0.1829          \\
crypto                      & \textbf{0.0001} & \textbf{0.0000} & \textbf{0.0000} & 0.9201          & \textbf{0.0030} & 0.3723          & 0.1607          & 0.2976          & 0.1012          & 0.3163          \\
ICO                         & 0.0800          & \textbf{0.0083} & 0.0156          & 0.1037          & 0.0814          & 0.0926          & 0.7811          & 0.7994          & 0.6019          & 0.4251          \\
coin market                 & \textbf{0.0013} & \textbf{0.0023} & \textbf{0.0002} & 0.4380          & \textbf{0.0001} & 0.7448          & 0.5298          & 0.4680          & 0.3202          & 0.5332          \\
soft fork                   & 0.0705          & 0.0147          & 0.0146          & 0.0546          & 0.3695          & 0.5942          & 0.6413          & 0.5173          & 0.1524          & 0.0963          \\
block producer              & 0.0186          & 0.0692          & 0.0966          & 0.4160          & 0.7026          & 0.4084          & \textbf{0.0012} & 0.2283          & 0.2444          & 0.5688          \\
distributed ledger          & \textbf{0.0000} & \textbf{0.0003} & \textbf{0.0006} & 0.1367          & 0.5447          & \textbf{0.0096} & 0.0598          & 0.0207          & 0.6033          & 0.1582          \\
digital fiat                & 0.3668          & 0.1305          & 0.2920          & 0.7814          & 0.1757          & \textbf{0.0035} & 0.1034          & 0.1989          & 0.3505          & 0.9097          \\
consensus                   & 0.1484          & 0.0414          & 0.1733          & 0.8250          & 0.2027          & 0.2676          & 0.3028          & 0.9427          & \textbf{0.0039} & 0.7434          \\
BTC-USD\_vol                & NA              & \textbf{0.0016} & 0.0169          & 0.4498          & 0.0976          & 0.5810          & 0.4332          & 0.3715          & 0.1185          & 0.3966          \\
ETH-USD\_vol                & 0.0211          & NA              & 0.6881          & 0.9887          & 0.7813          & 0.2604          & 0.3309          & 0.1331          & 0.2320          & 0.8606          \\
USDT-USD\_vol               & 0.0236          & 0.2817          & NA              & 0.9615          & 0.7748          & 0.4356          & 0.4442          & 0.1210          & 0.6740          & 0.9159          \\
BNB-USD\_vol                & 0.0407          & \textbf{0.0067} & \textbf{0.0001} & NA              & 0.0133          & 0.7227          & \textbf{0.0064} & 0.6081          & 0.3888          & 0.8414          \\
LTC-USD\_vol                & 0.0142          & \textbf{0.0008} & 0.0585          & \textbf{0.0011} & NA              & 0.8935          & 0.0882          & 0.5584          & 0.6791          & 0.8767          \\
ENJ-USD\_vol                & 0.0494          & 0.0905          & 0.0179          & 0.1367          & 0.6793          & NA              & 0.4163          & 0.5028          & 0.7517          & 0.6452          \\
ZEN-USD\_vol                & 0.2776          & 0.1236          & 0.0562          & 0.4521          & 0.1355          & 0.3684          & NA              & 0.2235          & 0.2593          & 0.0891          \\
NMC-USD\_vol                & 0.1121          & 0.0383          & 0.4738          & 0.8830          & 0.0831          & 0.2012          & 0.1662          & NA              & 0.6787          & 0.5602          \\
PPC-USD\_vol                & 0.9280          & 0.4225          & 0.3448          & 0.0272          & 0.3291          & 0.7256          & 0.6997          & 0.4033          & NA              & 0.9909          \\
FTC-USD\_vol                & 0.0690          & 0.1190          & 0.1657          & 0.5200          & 0.0164          & 0.5960          & 0.4863          & 0.0535          & 0.1750          & NA              \\ \hline
\end{tabular}
\end{adjustbox}
\end{table}

\end{document}